\documentclass[prd,superscriptaddress,unsortedaddress,twocolumn,showpacs,preprintnumbers,amsmath,amssymb]{revtex4}
\usepackage[dvips]{graphicx}
\usepackage{amsmath,amssymb,times}

\newcommand{\bequ}{\begin{equation}}
\newcommand{\eequ}{\end{equation}}
\newcommand{\bea}{\begin{eqnarray}}
\newcommand{\eea}{\end{eqnarray}}

\newcommand{\bfis}[1]{\mbox{\boldmath ${\scriptstyle #1}$}}

\newcommand{\vii}{{\bfis i}}

\newcommand{\vix}{{\bfis x}}

\DeclareSymbolFont{boldletters}{OML}{cmm} {b}{it}
\DeclareSymbolFontAlphabet{\mathbit}{boldletters}
\DeclareMathSymbol{\alpha}{\mathalpha}{letters}{"0B}
\DeclareMathSymbol{\beta}{\mathalpha}{letters}{"0C}
\DeclareMathSymbol{\gamma}{\mathalpha}{letters}{"0D}
\DeclareMathSymbol{\delta}{\mathalpha}{letters}{"0E}
\DeclareMathSymbol{\epsilon}{\mathalpha}{letters}{"0F}
\DeclareMathSymbol{\zeta}{\mathalpha}{letters}{"10}
\DeclareMathSymbol{\eta}{\mathalpha}{letters}{"11}
\DeclareMathSymbol{\theta}{\mathalpha}{letters}{"12}
\DeclareMathSymbol{\iota}{\mathalpha}{letters}{"13}
\DeclareMathSymbol{\kappa}{\mathalpha}{letters}{"14}
\DeclareMathSymbol{\lambda}{\mathalpha}{letters}{"15}
\DeclareMathSymbol{\mu}{\mathalpha}{letters}{"16}
\DeclareMathSymbol{\nu}{\mathalpha}{letters}{"17}
\DeclareMathSymbol{\xi}{\mathalpha}{letters}{"18}
\DeclareMathSymbol{\pi}{\mathalpha}{letters}{"19}
\DeclareMathSymbol{\rho}{\mathalpha}{letters}{"1A}
\DeclareMathSymbol{\sigma}{\mathalpha}{letters}{"1B}
\DeclareMathSymbol{\tau}{\mathalpha}{letters}{"1C}
\DeclareMathSymbol{\upsilon}{\mathalpha}{letters}{"1D}
\DeclareMathSymbol{\phi}{\mathalpha}{letters}{"1E}
\DeclareMathSymbol{\chi}{\mathalpha}{letters}{"1F}
\DeclareMathSymbol{\psi}{\mathalpha}{letters}{"20}
\DeclareMathSymbol{\omega}{\mathalpha}{letters}{"21}
\DeclareMathSymbol{\varepsilon}{\mathalpha}{letters}{"22}
\DeclareMathSymbol{\vartheta}{\mathalpha}{letters}{"23}
\DeclareMathSymbol{\varpi}{\mathalpha}{letters}{"24}
\DeclareMathSymbol{\varrho}{\mathalpha}{letters}{"25}
\DeclareMathSymbol{\varsigma}{\mathalpha}{letters}{"26}
\DeclareMathSymbol{\varphi}{\mathalpha}{letters}{"27}
\DeclareMathSymbol{\Gamma}{\mathalpha}{letters}{"00}
\DeclareMathSymbol{\Delta}{\mathalpha}{letters}{"01}
\DeclareMathSymbol{\Theta}{\mathalpha}{letters}{"02}
\DeclareMathSymbol{\Lambda}{\mathalpha}{letters}{"03}
\DeclareMathSymbol{\Xi}{\mathalpha}{letters}{"04}
\DeclareMathSymbol{\Pi}{\mathalpha}{letters}{"05}
\DeclareMathSymbol{\Sigma}{\mathalpha}{letters}{"06}
\DeclareMathSymbol{\Upsilon}{\mathalpha}{letters}{"07}
\DeclareMathSymbol{\Phi}{\mathalpha}{letters}{"08}
\DeclareMathSymbol{\Psi}{\mathalpha}{letters}{"09}
\DeclareMathSymbol{\Omega}{\mathalpha}{letters}{"0A}

\begin{document}
\preprint{SAGA-HE-287}
\title{Sign problem in $Z_3$-symmetric effective Polyakov-line model}

\author{Takehiro Hirakida}
\email[]{hirakida@email.phys.kyushu-u.ac.jp}
\affiliation{Department of Physics, Graduate School of Sciences, Kyushu University,
             Fukuoka 819-0395, Japan}
             
\author{Junpei Sugano}
\email[]{sugano@phys.kyushu-u.ac.jp}
\affiliation{Department of Physics, Graduate School of Sciences, Kyushu University,
             Fukuoka 819-0395, Japan}             
             
\author{Hiroaki Kouno}
\email[]{kounoh@cc.saga-u.ac.jp}
\affiliation{Department of Physics, Saga University,
             Saga 840-8502, Japan}

\author{Junichi Takahashi}
\email[]{j.t.mhjkk.f.c@gmail.com}
\affiliation{Division of Observation, Fukuoka Aviation Weather Station,
             Japan Meteorological Agency, Fukuoka 812-0005, Japan}

\author{Masanobu Yahiro}
\email[]{yahiro@phys.kyushu-u.ac.jp}
\affiliation{Department of Physics, Graduate School of Sciences, Kyushu University,
             Fukuoka 819-0395, Japan}

\date{\today}

\begin{abstract} 
As an effective model corresponding to $Z_3$-symmetric QCD ($Z_3$-QCD), 
we construct a $Z_3$-symmetric effective Polyakov-line model
($Z_3$-EPLM) by using the logarithmic fermion effective action. 
Since $Z_3$-QCD tends to QCD in the zero-temperature limit, 
$Z_3$-EPLM also agrees with the ordinary effective Polyakov-line model (EPLM) there; note that (ordinary) EPLM does not possess $Z_3$ symmetry. 
Our main purpose is to discuss a sign problem appearing in $Z_3$-EPLM. 
The action of $Z_3$-EPLM is real, when the Polyakov line is not only 
real but also its $Z_3$ images. 
This suggests that the sign problem becomes milder in $Z_3$-EPLM than in 
EPLM. In order to confirm this suggestion, we do lattice simulations 
for both EPLM and $Z_3$-EPLM 
by using the reweighting method with the phase quenched approximation. 
In the low-temperature region,
the sign problem is milder in $Z_3$-EPLM than in EPLM. 
We also propose a new reweighting method.  
This makes the sign problem very weak in $Z_3$-EPLM. 
\end{abstract}

\pacs{05.50.+q, 12.38.Aw, 25.75.Nq}
\maketitle

%%%%%%%%%%%%%%%%%%%%%%%%%%%%%%%%%%%%%%%%%%%%%%%%%%%%%%%%%%%%%%%%%%%%%%%%%%%
%%%%%  Introduction 
%%%%%%%%%%%%%%%%%%%%%%%%%%%%%%%%%%%%%%%%%%%%%%%%%%%%%%%%%%%%%%%%%%%%%%%%%%%
\section{Introduction}
\label{sec:intro}
%%%%%%%%%%%%%%%%%%%%%%%%%%%%%%%%%%%%%%%

Exploration of the QCD phase diagram at finite temperature $T$
and quark-number chemical potential $\mu$ is one of the most challenging
subjects in particle and nuclear physics, as well as in cosmology and astrophysics. 
For $\mu =0$, lattice QCD (LQCD)
simulations as a first-principle calculation are well established and 
yield much knowledge on hot-QCD matter.
For finite $\mu$, however, 
there is well-known difficulty in LQCD; namely, the so-called sign problem. 
At finite $\mu$, effective action $S_{\rm eff}$,
which is obtained after the  
integration of the quark fields in the grand canonical
partition function, is complex in general, and we cannot regard $e^{-S_{\rm eff}}$ as the probability function
that determines the realization probability of gauge configurations. 
This makes the importance sampling method impractical.   
To evade the sign problem, several approaches, e.g., the reweighting method~\cite{Fodor},
the Taylor expansion method~\cite{Allton,Ejiri_density}
and the analytic continuation from imaginary $\mu$
to real $\mu$~\cite{FP,D'Elia,D'Elia3,FP2010,Nagata,Takahashi}
were used. 
Recently, great progresses were made by the complex Langevin simulation
\cite{Aarts_CLE_1,Aarts_CLE_2,Aarts_CLE_3,Sexty,Aarts_James,Greensite:2014cxa}
and the Picard-Lefschetz thimble theory
\cite{Aurora_thimbles,Fujii_thimbles,Tanizaki,Tanizaki_2}. 
However, particularly for the region of $ \mu/T >1$ in $\mu$--$T$ 
plane, 
our understanding of the QCD phase diagram is still far from perfection. 

It is expected that $Z_3$ symmetry plays an important role
in solving the sign problem. 
It was conjectured~\cite{Langfeld_center} that the center-dressed quark
undergoes a new phase with the Fermi-Einstein condensation in cold
and dense matter, 
and the phenomenon is  a key to
the solution of the Silver Blaze problem~\cite{Barbour,Cohen}. 
It was shown that, by using the properties of $Z_3$ group,
an effective center field theory with the sign problem can be transformed
into a flux model with no sign problem~\cite{Condella,Z3C}. 
In Ref.~\cite{Z3A}, it was suggested that the sign problem in full QCD
with no exact symmetry may be cured to some extent
by using $Z_3$-averaged subset method. 
However, these methods are not adequate to solve the sign problem in QCD completely.      
 
In the pure SU(3) gauge theory, $Z_3$ symmetry is exact
and governs the confinement-deconfinement transition.
The Polyakov-line (loop)~\cite{Polyakov} is defined
as an exact order parameter
of the confinement-deconfinement transition. 
However, in full QCD with dynamical quarks,
$Z_3$ symmetry is not exact anymore, 
and it is not trivial that the Polyakov line is an order parameter
for the confinement-deconfinement transition~\cite{miyahara2}.  
In order to study the relation between the confinement-deconfinement transition
and the Polyakov line,
$Z_3$-symmetric QCD-like theory was proposed
in Refs.~\cite{Kouno_TBC,Sakai_TBC,Kouno_adjoint,Kouno_TBC_2,Kouno_DFTBC}.
We call the theory $Z_3$-QCD in this paper. 
$Z_3$-QCD was studied at first by 
the effective model~\cite{Kouno_TBC,Sakai_TBC,Kouno_adjoint,Kouno_TBC_2,Kouno_DFTBC}, and QCD simulations were recently made for $\mu =0$~\cite{Iritani}.

It was conjectured~\cite{Kouno_DFTBC} that
the sign problem is milder in $Z_3$-QCD than in the ordinary QCD. 
The study of the sign problem in $Z_3$-QCD is important,
since $Z_3$-QCD tends to three-flavor QCD in the limit $T\to 0$. 
To examine the conjecture,
Hirakida \textit{et al.}~\cite{Hirakida} constructed
a $Z_3$-symmetric 3D Potts model as a toy model of $Z_3$-QCD,
and studied a sign problem appearing in the model. 
It was found that the sign problem in the $Z_3$-symmetric 3D Potts model
is much milder than in the ordinary 3D 3-state Potts model~\cite{DeGrand,Karsch,Alford} with no $Z_3$ symmetry, 
even when several states 
are included in the calculations together with three $Z_3$-elements. 
However, the correspondence
between $Z_3$-QCD and the $Z_3$-symmetric 3D Potts model
defined in Ref.~\cite{Hirakida} is only qualitative. 

In this paper, we construct a $Z_3$-symmetric
effective Polyakov-line model ($Z_3$-EPLM).
The $Z_3$-EPLM is an effective model for 
$Z_3$-QCD in the heavy-quark and high-density limit,
and is closer to $Z_3$-QCD than the $Z_3$-symmetric 3D Potts model. 
First, we  show that the fermionic part of $Z_3$-EPLM action 
can be expressed by ``cubic Polyakov line'' 
that is invariant under the $Z_3$ transformation. 
Second, we find that 
the confinement state and 
the three deconfinement states based on $Z_3$ symmetry
are degenerate in $Z_3$-EPLM, 
when the pure-gauge contribution is neglected 
in the path integral. 
Doing lattice calculations based on the reweighting method 
with the phase quenched approximation, 
we conclude that, at low temperature, the sign problem is milder
in $Z_3$-EPLM than in EPLM with no $Z_3$ symmetry. 
Finally, we propose a new reweighting method. This method 
reduces the sign problem considerably. In particular, 
the problem becomes very weak in $Z_3$-EPLM. 

This paper is organized as follows. 
In Sec.~\ref{Z3QCD}, we review the formalism of $Z_3$-QCD. 
In Sec.~\ref{EPLM}, we formulate $Z_3$-EPLM,
and also discuss how important the particle-hole (P-H) symmetry~\cite{RF_PHS} is in EPLM. 
In Sec.~\ref{Section_RW}, the reweighting methods 
we use  are explained. 
Numerical simulations are done for the models in Sec.~\ref{numerical}.  
Section~\ref{summary} is devoted to a summary.

%%%%%%%%%%%%%%%%%%%%%%%%%%%%%%%%%%%%%%%%%%%%%%%%%%%%%%%%
%%%% effective Polyakov-line model
%%%%%%%%%%%%%%%%%%%%%%%%%%%%%%%%%%%%%%%%%%%%%%%%%%%%%%%%
\section{$Z_3$-QCD}
\label{Z3QCD}
%%%%%%%%%%%%%%%%%%%%%%%%%%%%%%%%%%%%%%%%%%%%%%%%%%%%%%%%

In this section, we review the formalism of $Z_3$-QCD~\cite{Kouno_TBC,Sakai_TBC,Kouno_adjoint,Kouno_TBC_2,Kouno_DFTBC,Iritani,Hirakida}.  
The grand canonical partition function of
three-flavor QCD with a common quark mass $m$,
quark-number chemical potential $\mu$ and temperature $T~(=1/\beta )$ is given by  
%%%%%%%%%%%%%%%%%%
\begin{eqnarray}
Z&=&\int {\cal D}A_\mu {\cal D}\bar{q}{\cal D}q e^{-S}; 
\label{Z_QCD}\\
S&=&S_{\rm G}+S_{\rm Q}
\nonumber\\
&=&\int_0^\beta dx_4\int_{-\infty}^\infty d^3\mathbf{x}{\cal L}_{\rm G}+\int_0^\beta dx_4\int_{-\infty}^\infty d^3\mathbf{x}{\cal L}_{\rm Q}; 
\label{S_QCD}\\
{\cal L}_{\rm G}&=&{1\over{4g^2}}{F_{\mu\nu}^a}^2 ={1\over{2g^2}}{\rm Tr}\left[ F_{\mu\nu}^2\right],  
\label{L_G}\\
{\cal L}_{\rm F}&=&\bar{q}{\cal M}q, 
\label{L_Q}
\end{eqnarray}
%%%%%%%%%%%%%%%%%  
where 
%%%%%%%%%%%%%%%%%
\begin{eqnarray}
{\cal M}&=&\gamma_\mu D_\mu +m-\mu\gamma_4, 
\label{calM}\\
F_{\mu\nu}&=&\partial_\mu A_\nu -\partial_\nu A_\mu +i[A_\mu ,A_\nu ], 
\label{Fmunu}\\
D_\mu &=&\partial_\mu -iA_\mu,~~~~~~A_\mu =g\sum_{a=1}^8 A_\mu^a{\lambda^a\over{2}}, 
\label{Dmu_Amu} 
\end{eqnarray}
%%%%%%%%%%%%%%%%%
with the quark field $q$, the gluon field $A_\mu^a$, the gauge coupling $g$
and the Gell-Mann matrices $\lambda^a$. 
The temporal anti-periodic boundary condition on the quark field is given by 
%%%%%%%%%%%%%%%%%
\begin{eqnarray}
q(x_4=\beta,{\bf x})=-q(x_4=0,{\bf x}), 
\label{boundary_q}
\end{eqnarray}
%%%%%%%%%%%%%%%%%
while the gluon field $A_\mu^a$ obeys the periodic boundary condition 
in the temporal direction. 
Note that $q$ is a vector also in flavor space. 
Hereafter, we denote each flavor component of $q$ as $q_f$.

The Lagrangian density ${\cal L}_{\rm G}+{\cal L}_{\rm F}$
is invariant under the $Z_3$ transformation, 
% {\bf Comment 1-1}
%%%%%%%%%%%%%%%%%
\begin{eqnarray}
q \to q^\prime =Uq,~~~A_\mu\to A_\mu^\prime =UA_\mu U^{-1}+i(\partial_\mu U)U^{-1}, 
\label{Z_3_transformation}
\end{eqnarray}
%%%%%%%%%%%%%%%%
where $U=\exp{(i\alpha_a (x_4,{\bf x})\lambda^a /2)}$ is an element of SU(3) group characterized by real functions $\alpha_a (x_4,{\bf x})$ satisfying the temporal boundary condition
%%%%%%%%%%%%%%%%
\begin{eqnarray}
U(x_4=\beta ,{\bf x})=e^{-i2\pi k/3}U(x_4=0,{\bf x})
\label{Z_3_twist}
\end{eqnarray}
%%%%%%%%%%%%%%%%
for any integer $k$. 
However, the condition (\ref{boundary_q}) on the quark field is changed by the $Z_3$ transformation (\ref{Z_3_transformation}) into
%%%%%%%%%%%%%%%%%
\begin{eqnarray}
q(x_4=\beta,{\bf x})=-e^{i2\pi k/3}q(x_4=0,{\bf x}). 
\label{boundary_q_Z3}
\end{eqnarray}
%%%%%%%%%%%%%%%%%
Thus, in full QCD with dynamical quarks,
$Z_3$ symmetry is explicitly broken through the quark boundary condition. 

To recover $Z_3$ symmetry,
one can consider 
the flavor dependent twist boundary condition (FTBC)~\cite{Kouno_TBC,Sakai_TBC,Kouno_adjoint,Kouno_TBC_2,Kouno_DFTBC},  
%%%%%%%%%%%%%%%%%%%%%%%%%%%%%%%%%
\begin{eqnarray}
q_f(x_4=\beta,{\bf x})=-e^{-i\theta_f}q_f(x_4=0,{\bf x}); 
\nonumber\\
\theta_f={2\pi\over{3}}f~~~~~(f=-1,0,1), 
\label{bound_FTBC}
\end{eqnarray}
%%%%%%%%%%%%%%%%%%%%%%%%%%%%%%%%% 
instead of the condition (\ref{boundary_q}). 
Here, the flavor indices are represented by the number $-1,0,1$ for convenience.
Under the $Z_3$ transformation (\ref{Z_3_transformation}), the FTBC (\ref{bound_FTBC}) is transformed into 
% {\bf Comment1-2}
%%%%%%%%%%%%%%%%%%%%%%%%%%%%%%%%%
\begin{eqnarray}
q_f(x_4=\beta,{\bf x})=-e^{-i\theta_f^\prime }q_f(x_4=0,{\bf x}); 
\nonumber\\
\theta_f^\prime ={2\pi\over{3}}(f-k)~~~~~(f=-1,0,1). 
\label{FTBC_Z3}
\end{eqnarray}
%%%%%%%%%%%%%%%%%%%%%%%%%%%%%%%%%
The transformed boundary condition (\ref{FTBC_Z3}) returns to the original one (\ref{bound_FTBC}) by relabeling the flavor indices $f-k$ as $f$. 
Hence, the QCD-like theory with the FTBC (\ref{bound_FTBC})
is invariant under the $Z_3$ transformation. 
In this paper, this theory is referred to as $Z_3$-QCD. 
It should be noted that $Z_3$-QCD agrees with
original QCD in the limit $T\to 0$,
since the boundary condition is not relevant in this limit.  

When the quark fields $q_f$ are transformed into \cite{RW}
%%%%%%%%%%%%%%%%%%%%%%%%%%%%%%%%%%
\begin{eqnarray}
q_f\to e^{-i\theta_fT x_4 }q_f, 
\label{RW_trans}
\end{eqnarray}
%%%%%%%%%%%%%%%
the FTBC (\ref{bound_FTBC}) returns to the ordinary anti-periodic
boundary condition (\ref{boundary_q}). 
However, the quark part $\mathcal{L}_{\rm F}$ of the Lagrangian density is changed into 
%%%%%%%%%%%%%%%%%%%%%%%%%%%%%%%%%
\begin{eqnarray}
{\cal L}^{\hat{\theta}}_{\rm F}
 =\bar{q}(\gamma_\mu D_\mu +m-i{\hat{\theta}}T\gamma_{4})q, 
\label{L_theta_hat}
\end{eqnarray}
%%%%%%%%%%%%%%
where 
%%%%%%%%%%%%%%%%%%%%%%%%%%%%%%%%%
\begin{eqnarray}
\hat{\theta}={\rm diag}(\theta_{-1},\theta_0,\theta_1)={\rm diag}(-2\pi /3,0,2\pi /3)
%\nonumber\\
\label{hat_theta}
\end{eqnarray}
%%%%%%%%%%%%%%
is the flavor-dependent imaginary chemical potential normalized by $T$.  
(In the pioneering work of Ref.~\cite{HT_canonical}, Hasenfratz and Toussaint introduced the fictitious flavor-dependent imaginary chemical potential to make the non-zero triality sectors of canonical partition function vanish more rigidly in LQCD simulations. ) 
The flavor-dependent imaginary chemical potential partially breaks
flavor SU(3) symmetry.
In the chiral limit $m\to 0$,
due to the existence of $\hat{\theta}$,
global ${\rm SU}_{\rm V}(3) \otimes {\rm SU}_{\rm A}(3)$ symmetry
is broken to $({\rm U}_{\rm V}(1))^2\otimes ({\rm U}_{\rm A}(1))^2$~\cite{Kouno_adjoint}. 
The remaining symmetry is broken into $({\rm U}_{\rm V}(1))^2$,
if chiral symmetry is spontaneously broken. 
Recently, the FTBC was also discussed in the context of spontaneous chiral symmetry breaking~\cite{Cherman,Liu}.

%%%%%%%%%%%%%%%%%%%%%%%%%%%%%%%%%%%%%%%%%%%%%%%%%%%%%%%%
%%%% effective Polyakov-line model
%%%%%%%%%%%%%%%%%%%%%%%%%%%%%%%%%%%%%%%%%%%%%%%%%%%%%%%%
\section{$Z_3$-symmetric effective Polyakov-line model}
\label{EPLM}
%%%%%%%%%%%%%%%%%%%%%%%%%%%%%%%%%%%%%%%%%%%%%%%%%%%%%%%%

\subsection{Effective Polyakov-line model}
\label{EPLM_A}

In this section, we formulate $Z_3$-symmetric EPLM and examine its properties. 
Before going to the discussion, we first review the ordinary EPLM.  
The grand canonical partition function of EPLM is given by~\cite{Aarts_James,Greensite:2014cxa}
%%%%%%%%%%%%%%
\begin{eqnarray}
Z&=&\int {\cal D}U \exp{\left(-S_{\rm F}-S_{\rm G}\right)}; 
\label{Z_EPL}
\\
S_{\rm F}&=& \sum_{\vix} {L}_{\rm F}({\bf x}), 
\label{SF_EPL}
\\
S_{\rm G}&=&-\kappa\sum_{\vix}\sum_{i =1}^3
\left( {\rm Tr}[U_{\vix}]{\rm Tr}[U_{\vix+ \vii}^\dagger]
+{\rm Tr}[U_{\vix}^\dagger]{\rm Tr}[U_{\vix+ \vii}]
\right),
\nonumber\\
\label{SG_EPL}
\end{eqnarray}
%%%%%%%%%%%%%
where $U_\vix$ is the Polyakov-line (loop) holonomy
and the symbol $\mathbf{i}$ is an unit vector for $i$-th direction. 
The site variable $x$ runs over a 3-dimensional lattice.

The constant parameter $\kappa$ is related to temperature $T=1/\beta$. 
Roughly speaking,
large (small) $\kappa$ corresponds to high (low)
temperature~\cite{DeGrand}, but the relation is not so simple. 
In this paper, we treat $\kappa$ just as a parameter independent of $T$,
while the other parameters 
with the energy dimension are always normalized by $T$.        
In Ref.~\cite{Greensite:2014cxa}, a more-complicated form with a lot of parameters is used for the gauge action $S_{\rm G}$, but we take 
a simple form as in Ref.~\cite{Aarts_James} in order to know qualitative 
properties of the phase structure and the sign problem of this model 
in a wide range of the parameter set.  
Using the temporal gauge, we parameterize $U_\vix$ as ~\cite{Greensite:2014cxa}
%%%%%%%%%%%%%%%%
\begin{eqnarray}
U_\vix &=&{\rm diag} \left( e^{i\varphi_{r,\vix}},e^{i\varphi_{g,\vix}},e^{i\varphi_{b,\vix}}\right), 
\label{U_x}
\\
U_\vix^\dagger &=&{\rm diag}\left( e^{-i\varphi_{r,\vix}},e^{-i\varphi_{g,\vix}},e^{-i\varphi_{b,\vix}}\right)
\label{U_x_dagger}
\end{eqnarray}
%%%%%%%%%%%%%%
with the condition $\varphi_{r,\vix}+\varphi_{g,\vix}+\varphi_{b,\vix}=0$, 
and define the (traced) Polyakov line (loop) $P_\vix$ and its conjugate $P_\vix^*$ as 
%%%%%%%%%%%%%%
\begin{eqnarray}
P_\vix ={1\over{3}}{\rm Tr}\left[ U_\vix \right]&=&{1\over{3}}\left( e^{i\varphi_{r,\vix}}+e^{i\varphi_{g,\vix}}+e^{i\varphi_{b,\vix}}\right), 
\label{P_x}
\\
P_\vix^* ={1\over{3}}{\rm Tr}\left[ U_\vix^\dagger \right]&=&{1\over{3}}\left( e^{-i\varphi_{r,\vix}}+e^{-i\varphi_{g,\vix}}+e^{-i\varphi_{b,\vix}}\right). 
\nonumber\\
\label{P_x_conj}
\end{eqnarray}
%%%%%%%%%%%%%%%
In this paper,  instead of $U_\vix$ and $U_\vix^\dagger$, we treat the phase variables $\varphi_{r,\vix}$ and $\varphi_{g,\vix}$ as dynamical variables.  
The Haar measure ${\cal D}U$ in the path integral (\ref{Z_EPL}) is rewritten into~\cite{Greensite:2014cxa}
%%%%%%%%%%%%%%%
\begin{eqnarray}
&&{\cal D}U=e^{-S_{\rm H}(\varphi_{r,\vix},\varphi_{g,\vix})}{\cal D}\varphi_{r,\vix}{\cal D}\varphi_{g,\vix};
\label{HaarM}\\
&&S_{\rm H}=\sum_{\vix} {L}_{\rm H}({\bf x})
\label{HaarS} , \\
&&{L}_{\rm H}({\bf x})=-\log{}\Bigl\{\sin^2{\left({\varphi_{r,\vix}-\varphi_{g,\vix}\over{2}}\right)}
\nonumber\\
&&\times \sin^2{\left({2\varphi_{r,\vix}+\varphi_{g,\vix}\over{2}}\right)}
\sin^2{\left({\varphi_{r,\vix}+2\varphi_{g,\vix}\over{2}}\right)
\Bigr\}}.
\label{HaarL}
\end{eqnarray}
%%%%%%%%%%%%%%
Note that, for simplicity of notation, we use dimensionless volume $V=N_s^3$ and dimensionless Lagrangian density $L$, where $N_s$ is the number of lattice sites in one spatial direction.
Hereafter, for simplicity, we refer to this three-flavor EPLM with no $Z_3$ symmetry as EPLMWO.
 
For the fermionic Lagrangian density, 
we consider a logarithmic one of Ref. \cite{Greensite:2014cxa}: 
%%%%%%%%%%%%%%
\begin{eqnarray}
{L}_{\rm F}[\mu,\varphi_{c,\vix}]
&=&-\log{\Bigl( {\rm det}\left[1+e^{\beta(\mu -M)}U_\vix\right]^{2N_f}}
\nonumber\\
&&
\times 
{\rm det}\left[1+e^{-\beta (\mu+M)}U_\vix^\dagger \right]^{2N_f}\Bigr)
\nonumber\\
&=&-2N_f\sum_{c=r,g,b}\Bigl\{\log{\Bigl(1+e^{\beta(\mu -M+i\varphi_{c,\vix})}\Bigr)}
\nonumber\\
&&
+\log{\Bigl(1+e^{-\beta(\mu +M+i\varphi_{c,\vix})}\Bigr)}\Bigr\}
\nonumber\\
&=&-2N_f\Bigl\{\log{\Bigl(1+3e^{\beta(\mu -M)}P_\vix }
\nonumber\\
&&+3e^{2\beta(\mu -M)}P_\vix^* +e^{3\beta(\mu -M)}\Bigr)
\nonumber\\
&&+\log{\Bigl(1+3e^{-\beta(\mu +M)}P_\vix^*}
\nonumber\\
&&
+3e^{-2\beta(\mu +M)}P_\vix +e^{-3\beta(\mu +M)}\Bigr)\Bigr\}, 
\label{log_L}   
\end{eqnarray}
%%%%%%%%%%%%%%
where $M$ is the quark mass. 

It is easily seen that, in the limit 
$M\to \infty$, ${L}_{\rm F}$ becomes real at $\mu =M$, since 
$e^{\beta (\mu-M)}=e^{2\beta (\mu-M)}=1$ and $e^{-\beta (\mu+M)}=e^{-2\beta (\mu+M)}=0$ in the last line of Eq. (\ref{log_L}). 
As seen below, the reality of ${L}_{\rm F}$
at $\mu =M$ is related to the particle-hole symmetry in EPLM.

%%%%%%%%%%%%%%%%%%%%%%%%%%%%%%%%%%%%%%%%%%%%%%%%%
\subsection{Particle-hole symmetry}
\label{subs_PH}
%%%%%%%%%%%%%%%%%%%%%%%%%%%%%%%%%%%%%%%%%%%%%%%%%   

The particle contribution in ${L}_{\rm F}$ can be rewritten into 
%%%%%%%%%%%%%%%%%%%%%%%%%%%%%%%%%%%%%%%%%
\begin{eqnarray}
&&{L}_{{\rm F,p}}[\mu =M-\Delta \mu, \varphi_{c,\vix}]
\nonumber\\
&&=-2N_f\sum_{c=r,g,b}\log{\left( 1+e^{\beta (-\Delta \mu +i\varphi_{c,\vix})}\right)}
\nonumber\\
&&=-2N_f\sum_{c=r,g,b}\Bigl\{\beta(-\Delta \mu +i\varphi_{c,\vix})
\nonumber\\
&&+\log{\left( 1+e^{\beta (\Delta \mu-i\varphi_{c,\vix})}\right)}\Bigr\}
\nonumber\\
&&=6N_f\beta \Delta \mu 
-2N_f\sum_{c=r,g,b}\log{\left( 1+e^{\beta (\Delta \mu-i\varphi_{c,\vix})}\right)}
\nonumber\\
&&=6N_f\beta \Delta \mu + {L}_{{\rm F,p}}[\mu =M+ \Delta \mu, -\varphi_{c,\vix}]. 
\label{p-h}
\end{eqnarray}
%%%%%%%%%%%%%%
The term $6N_f\beta\Delta \mu$ in the last line of (\ref{p-h}) does
not depend on the dynamical variable $\varphi_{c,\vix}$
and does not contribute to the expectation value of physical quantities.
Hence, the relation
${L}_{\rm F,p}[\mu =M-\Delta \mu, \varphi_{c,\vix}]={L}_{\rm F,p}[\mu =M+\Delta \mu,-\varphi_{c,\vix}]$ is effectively satisfied. 
Furthermore, since the antiparticle contribution in ${L}_{\rm F}$
is negligible in the limit $M\to \infty$,
we obtain $\langle \bar{P} (\mu =M-\Delta \mu )\rangle=\langle \bar{P}^*(\mu =M+\Delta \mu)\rangle$ and $\langle \bar{P}^*(\mu = M-\Delta \mu)\rangle=\langle \bar{P} (\mu =M+\Delta \mu )\rangle$. 
We have a relation $\langle O(\mu = M-\Delta \mu )\rangle=\langle O(\mu  =M+\Delta \mu )\rangle$ for  any quantity $O$ that 
does not depend on the sign of $\varphi_{c,\vix}$
and has no explicit $\mu$ dependence. 
This relation is nothing but the particle-hole (P-H) symmetry~\cite{RF_PHS}. 
From this symmetry, one can easily derive the relation
${L}_{\rm F}[\mu =M, \varphi_{c,\vix}]={L}_{\rm F,p}[\mu =M,-\varphi_{c,\vix}]$ in the limit $M\to\infty$. 
This relation ensures that ${L}_{\rm F}$ is real at $\mu =M$.  

It should be remarked that the effects of spatial momenta of quarks make 
the P-H symmetry invisible. 
In fact, in the Polyakov-loop extended Nambu--Jona-Lasinio (PNJL) model~\cite{Meisinger,Dumitru,Fukushima,Ratti,Megias}, the quark-loop contribution $\Omega_{\rm PNJL,F}$ 
of the thermodynamical potential density is given, 
under the mean field approximation, as  
%%%%%%%%%%%%%%%%%%%%%%%%%%%%%%%%%%%%%%%%%
\begin{eqnarray}
&&\Omega_{\rm PNJL,F}=-2N_f\sum_{c=r,g,b}\int {dp\over{2\pi^2}}\Bigl\{E_p
\nonumber\\
&&+T\log{\left( 1+e^{\beta (\mu - E_p)}e^{i\phi_{c}}\right)}
 \nonumber\\
&&+T\log{\left( 1+e^{-\beta(\mu + E_p)}e^{-i\phi_{c}}\right)}
\Bigr\}, 
\nonumber\\
\label{PNJL}
\end{eqnarray}
%%%%%%%%%%%%%%
where $E_p=\sqrt{p^2+M^2}$, $p$ is the absolute value of quark spatial momentum, and $\phi_c$ is similar to $\varphi_{c,\vix}$ but does not depend on the spatial coordinate ${\bf x}$. 
Up to the factor $\mu -E_p$ that does not depend on $\phi_c$,
the logarithmic function $\log{\left( 1+e^{\beta (\mu - E_p)}e^{i\phi_{c}}\right)}$
 is symmetric under the transformation, $\mu =E_p-\Delta \mu \to \mu =E_p+\Delta \mu$ and $\phi_c\to -\phi_c$, but the location of the symmetric point $\mu =E_p$ depends on $p$.
Hence, the symmetry is invisible when the integration over $p$ is performed.  
The symmetry does not appear explicitly in QCD where quarks 
have spatial momenta. 
Therefore, the appearance of the {\it explicit} P-H symmetry indicates the limitation of EPLM. 
The EPLM is considered to be valid as an effective model of QCD only in the region where $\mu$ is not much larger than $M$. 

%%%%%%%%%%%%%%%%%%%%%%%%%%%%%%%%%%%%%%%%%%%%%%%%%%%%%%%%%%
\subsection{$Z_3$-symmetric effective Polyakov-line model ($Z_3$-EPLM)}
\label{subs_Z3_EPLM} 
%%%%%%%%%%%%%%%%%%%%%%%%%%%%

Since the traced Polyakov line is not invariant under the $Z_3$ transformation,
the Lagrangian density (\ref{log_L}) is not invariant under the $Z_3$ transformation. 
To preserve $Z_3$ symmetry, we consider the three-flavor case
and introduce the flavor-dependent imaginary chemical potential $i\theta_fT~(f=u,d,s)$,
where $(\theta_u,\theta_d,\theta_s)=(2\pi /3,-2\pi/3,0)$.   
The corresponding Lagrangian density is given by 
%%%%%%%%%%%%%%
\begin{eqnarray}
&&{{L}}_{{\rm F},Z_3}[\mu,\varphi_{c,\vix}]
\nonumber\\
&&=-2\sum_{f=u,d,s}\sum_{c=r,g,b}\Bigl\{\log{\left(1+e^{\beta(\mu -M+i\theta_f+i\varphi_{c,\vix})}\right)}
\nonumber\\
&&+\log{\left( 1+e^{-\beta(\mu +M+i\theta_f+i\varphi_{c,\vix})}\right)}\Bigl\}
\nonumber
\end{eqnarray}
%%%%%%%%%%%%%%%%
%%%%%%%%%%%%%
\begin{eqnarray}
~~~~~~~~&=&-2\sum_{c=r,b,c}\Bigl\{\log{\left( 1+e^{3\beta (\mu -M+i\varphi_{c,\vix} )} \right)}
\nonumber\\
&&+
\log{\left(  1+e^{-3\beta (\mu +M+i\varphi_{c,\vix} )}\right)}\Bigr\}
\nonumber\\
&=&-2\log{\Bigl(1+3e^{3\beta(\mu -M)}Q_\vix}
\nonumber\\
&&+3e^{6\beta(\mu -M)}Q_\vix^* 
+e^{9\beta(\mu -M)}\Bigr)
\nonumber\\
&&-2\log{\Bigl(1+3e^{-3\beta(\mu +M)}Q_\vix^*}
\nonumber\\
&&+3e^{-6\beta(\mu +M)}Q_\vix 
+e^{-9\beta(\mu +M)}\Bigr), 
\label{log_L_Z3}   
\end{eqnarray}
%%%%%%%%%%%%%%
where "(traced) cubic Polyakov-line" $Q_\vix$ and its conjugate $Q_\vix^*$ are defined by 
%%%%%%%%%%%%%%
\begin{eqnarray}
Q_\vix &=&{1\over{3}}{\rm Tr}\left[ (U_\vix)^3 \right]={1\over{3}}\left( e^{i3\varphi_{r,\vix}}+e^{i3\varphi_{g,\vix}}+e^{i3\varphi_{b,\vix}}\right), 
\label{L_x}
\\
Q_\vix^* &=& {1\over{3}}{\rm Tr}\left[ (U_\vix^\dagger )^3 \right]={1\over{3}}\left( e^{-3i\varphi_{r,\vix}}+e^{-3i\varphi_{g,\vix}}+e^{-3i\varphi_{b,\vix}}\right).
\nonumber\\
\label{L_x_conj}
\end{eqnarray}
%%%%%%%%%%%%%%%
It is easy to derive 
%%%%%%%%%%%%%%%
\begin{eqnarray}
Q_\vix &=&9\left\{(P_\vix)^3-P_\vix P_\vix^*\right\}+1,
 \label{L-P1}
\\
Q_\vix^* &=&9\left\{(P_\vix^*)^3-P_\vix P_\vix^*\right\} +1. 
\label{L-P2}
\end{eqnarray}
%%%%%%%%%%%%%%%
From (\ref{L-P1}) and~(\ref{L-P2}),
it is clear that, in $L_{{\rm F},Z_3}$,
the degeneration between
the ``deconfinement'' gauge state with $P_\vix =1$ and
the ``confinement'' state with $P_\vix =0$~\cite{Kouno_adjoint} occurs,
since $Q_\vix =Q_\vix^* =1$ in both the states.  
Note that the cubic Polyakov line
is not an order parameter of the confinement-deconfinement transition, since it is invariant under the $Z_3$-transformation. 
This property of $Q_\vix$ resembles
that of the Polyakov line $P_\vix^{\rm adj}$ in the adjoint representation~\cite{Kouno_adjoint}. 
However, there is an essential difference between them. 
In fact, $P_\vix^{\rm adj}$ is related to $P_\vix$ as  
%%%%%%%%%%%%%%%
\begin{eqnarray}
P_\vix^{\rm adj} &=&{1\over{8}}\left( 9P_\vix P_\vix^*-1\right). 
\label{L-adj}
\end{eqnarray}
%%%%%%%%%%%%%%%
Hence, it is $Z_3$-invariant and real. 
However, $Q_\vix$ is $Z_3$-invariant but not real in general. 
The sign problem exists in $Z_3$-EPLM, since 
${\rm Im}\left[Q_{\vix}\right]$ can be finite. 
We also remark that, as in the case of the ordinary EPLM, 
$Z_3$-EPLM has the explicit P-H symmetry
that does not appear in $Z_3$-QCD itself.
Hence $Z_3$-EPLM is also valid as an effective model of $Z_3$-QCD
only in the region where $\mu$ is not much larger than $M$.

%%%%%%%%%%%%%%%%%%%%%%%%%%%%
\subsection{Fermion potential in EPLM}
\label{EPLM0}
%%%%%%%%%%%%%%%%%%%%%%%%%%%%
In this subsection, we examine properties of the fermion potential $L_{\rm F}$ in EPLM.  
Note that the Lagrangian density or the potential is  dimensionless in our definition.

\begin{figure}[htbp]
\begin{center}
\includegraphics[width=0.35\textwidth,angle=-90]{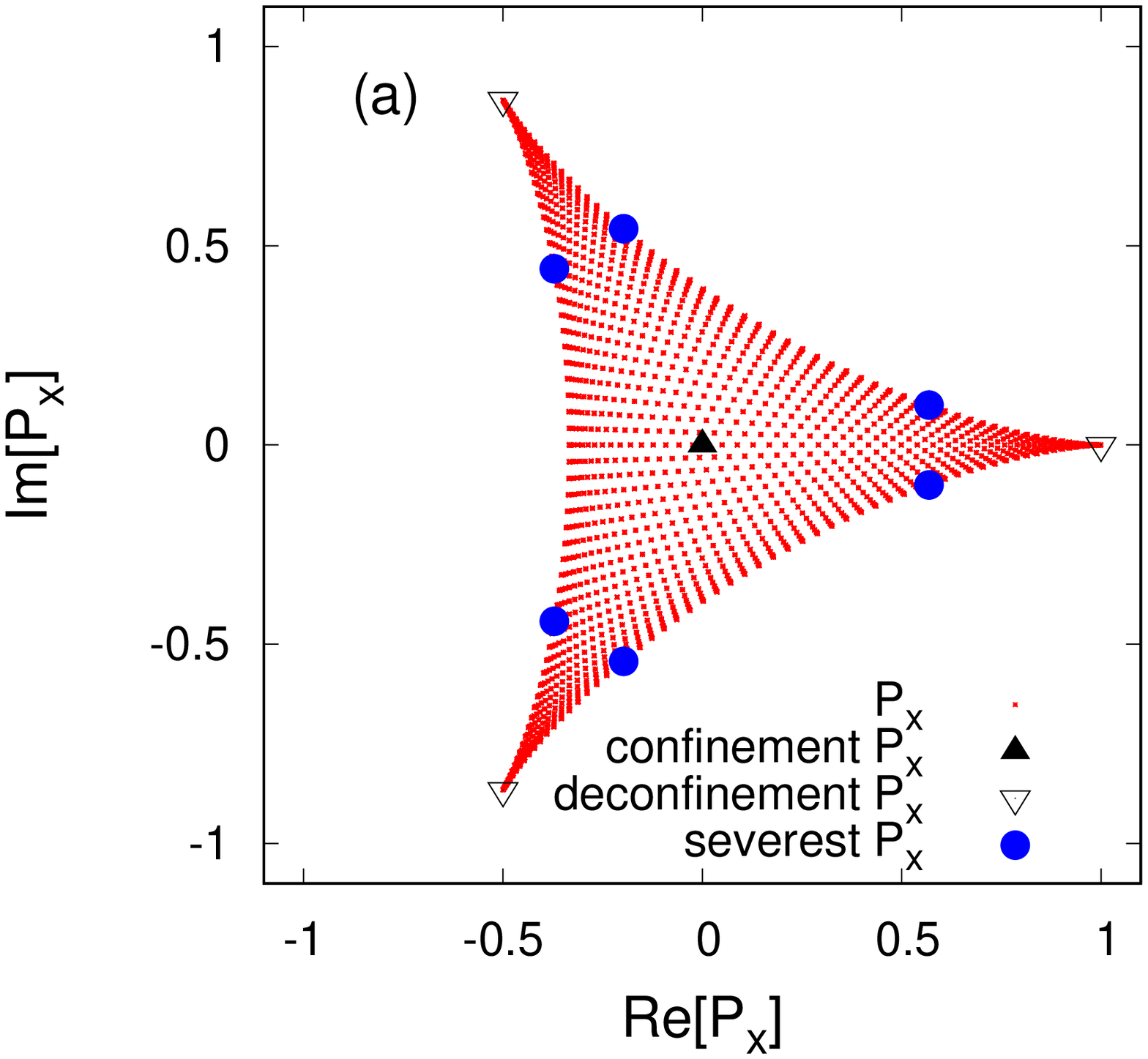}
\includegraphics[width=0.35\textwidth,angle=-90]{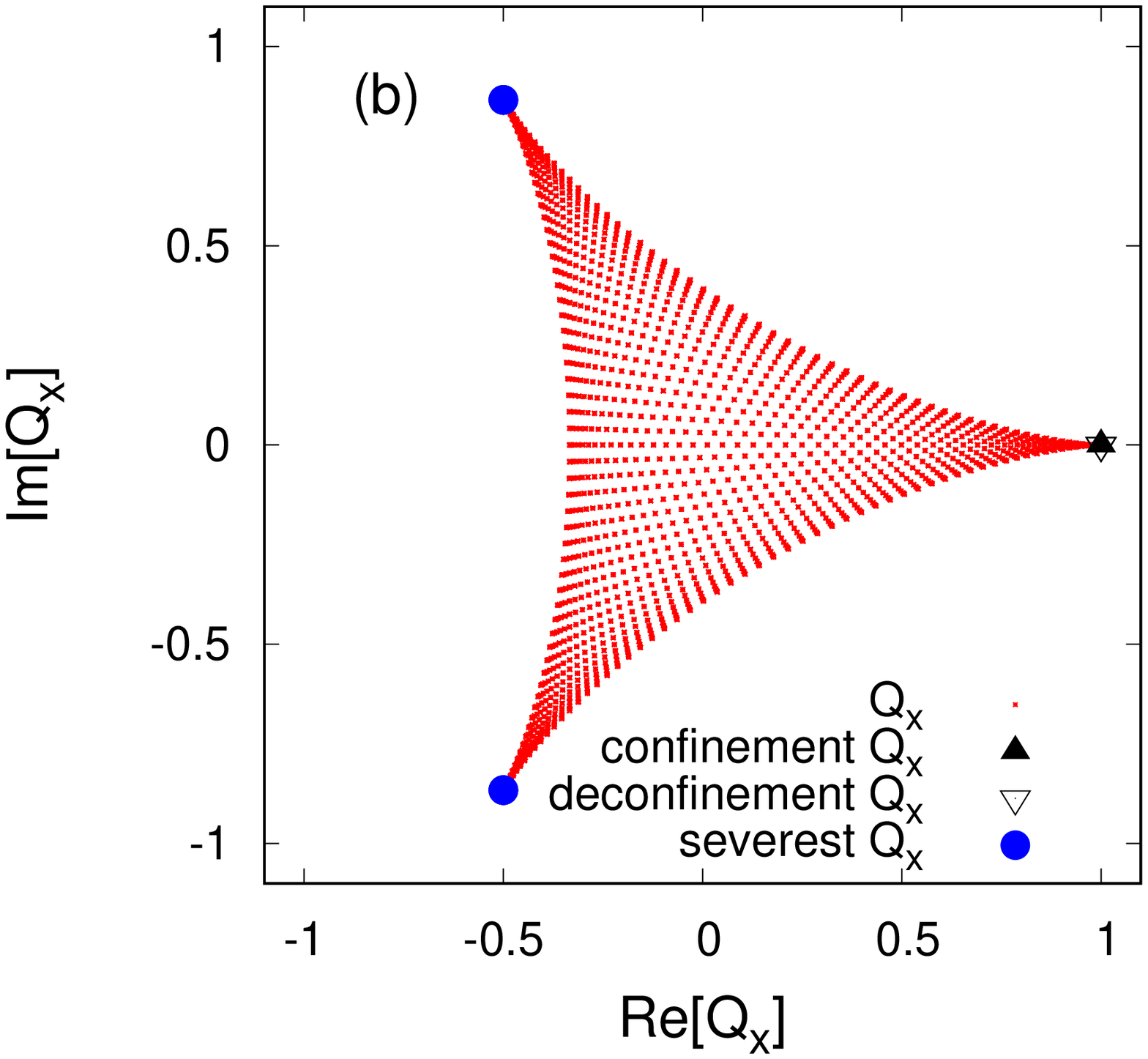}
\end{center}
\caption{Allowed regions of (a) $P_\vix$ and (b) $Q_\vix$ in the complex plane. 
In (a), three vertices correspond to the deconfinement points $(1,0)$, $(-1/2,\sqrt{3}/2)$ and $(-1/2,-\sqrt{3}/2)$, while they are degenerate at the point (1,0) in (b). 
In (b), the confinement point is also degenerate with the three deconfinement points.    
In $Z_3$-EPLM, the sign problem is severest at the blue circles. 
}
\label{Fig_P_and_L}
\end{figure}

In Fig.~\ref{Fig_P_and_L}, the values of $P_\vix$ and $Q_\vix$ are shown in their complex planes. 
At first glance, $Q_\vix$ has the same structure as $P_\vix$ in the  complex plane.  
However, there is an essential difference between them. 
In the complex plane of $P_\vix$, points $(-1/2,\pm \sqrt{3}/2)$ are 
$Z_3$-images of a deconfinement state $P_\vix =1=(1,0)$. 
However, in the complex plane of $Q_\vix$, points $(-1/2,\pm \sqrt{3}/2)$ are 
not $Z_3$-images of $P_\vix =1$. 
The $Z_3$-images of $P_\vix =1$ are degenerate with $P_\vix =1$ that 
corresponds to point $(1,0)$ in the complex plane of $Q_\vix$. 
  
Furthermore, as already mentioned in the previous subsection, the confinement state $P_\vix =0$ is also degenerate with $P_\vix =1$ corresponding 
to point $(1,0)$ in the complex plane of $Q_\vix$. 
More generally, the states $P_\vix =be^{\pm i2\pi/3}~~(b=-1/3\sim 1)$ 
are degenerate with $P\vix =b$  
corresponding to the real axis in the complex plane of $Q_\vix$. 

%%%%%%%%%%%%%%%%%%%%%%
%%%%%%%%%%%%

In the complex $Q_\vix$  plane, 
the origin corresponds to a configuration $(\varphi_{r,\vix},\varphi_{g,\vix},\varphi_{b,\vix})=(2\pi /9,-2\pi /9, 0)$ and its Weyl symmetry transformations, 
and points $(-1/2,\pm \sqrt{3}/2)$ correspond to 
configurations  $(\varphi_{r,\vix},\varphi_{g,\vix},\varphi_{b,\vix})=(\pm 2\pi /9,\pm 2\pi /9, \mp 4\pi /9)$ and their $Z_3$ and/or 
Weyl symmetry transformations. 
The latter points are denoted by the solid circles in Fig.~\ref{Fig_P_and_L}.   
At these points, the absolute value of ${\rm Im}[Q_\vix ]$ becomes maximum and the sign problem is severest in $Z_3$-EPLM when these states are favored.  

Since $P_\vix $ can be complex in general, 
the reality of the action is not ensured at finite $\mu$ and a sign problem occurs in EPLM.   
However, for simplicity, we ignore the effects of imaginary part ${\rm Im}[L_{\rm F}]$ (or ${\rm Im}[L_{{\rm F},Z_3}]$) in this subsection. 
If the sign problem is serious, 
the following discussion may not be applicable. 
We will discuss the sign problem in EPLM and $Z_3$-EPLM in 
Sec. \ref{EPLM0_sign}.    

\begin{figure}[htbp]
\begin{center}
\includegraphics[width=0.4\textwidth]{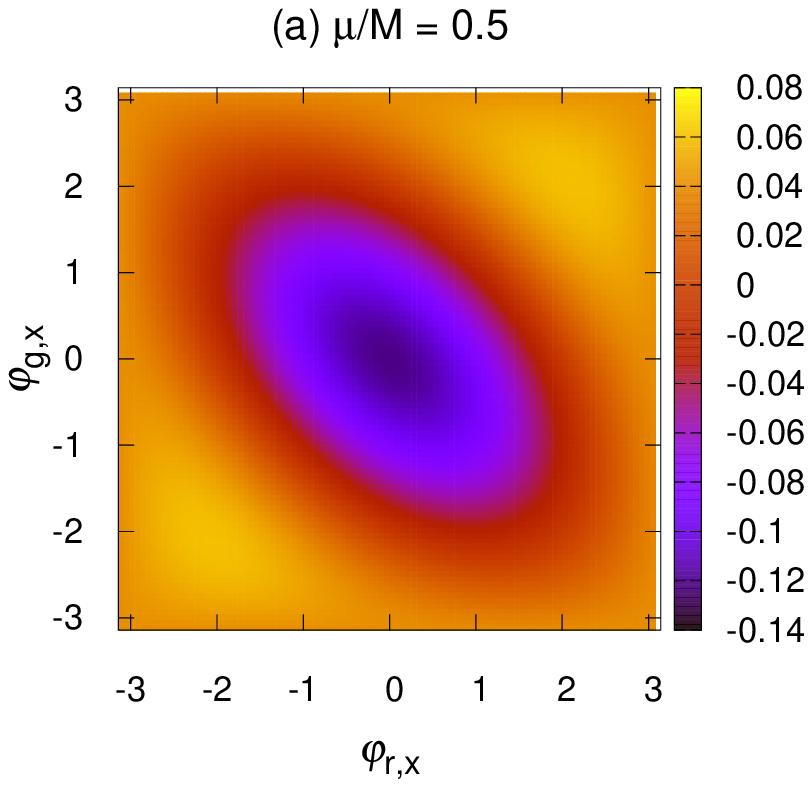}
\includegraphics[width=0.4\textwidth]{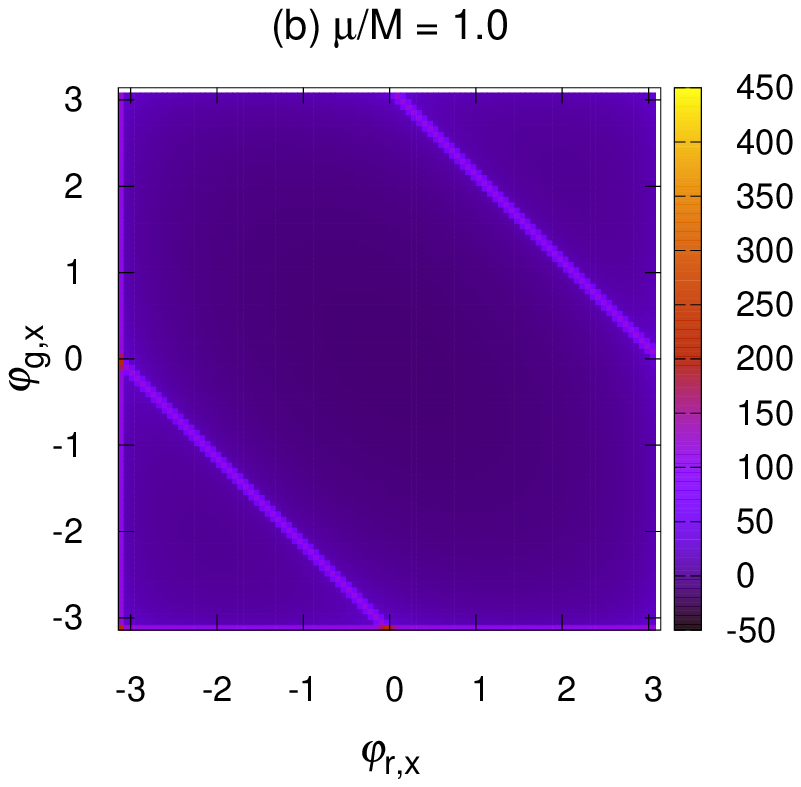}
\includegraphics[width=0.4\textwidth]{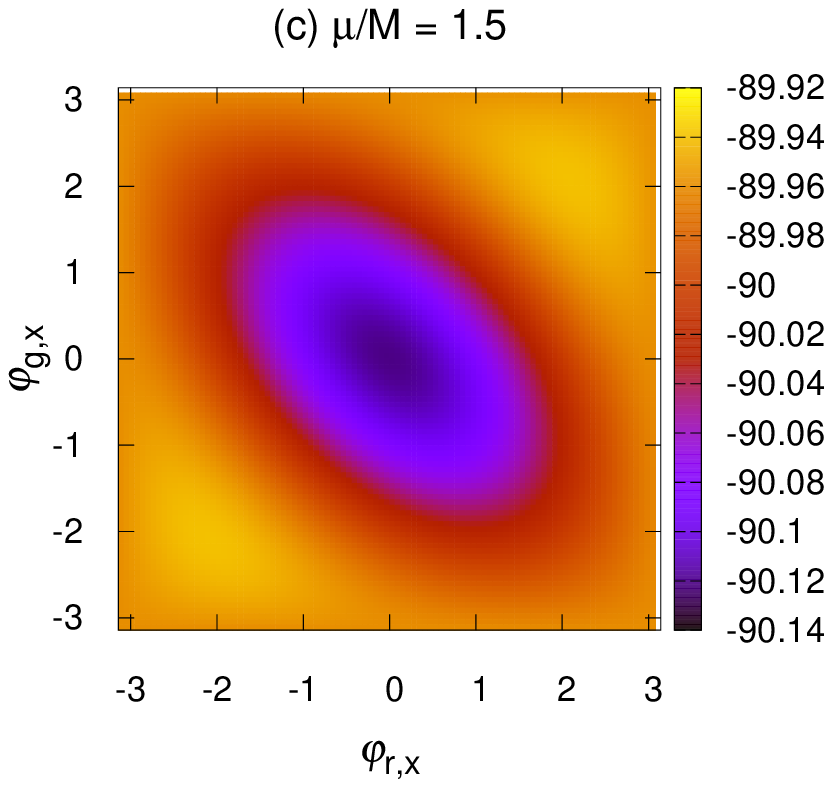}
\end{center}
\caption{${\rm Re}[{L}_{\rm F}]$ in $\varphi_{r,\vix}$--$\varphi_{g,\vix}$ plane for the case of EPLMWO.%when EPLMWO is used.  
The fermion potential takes minimum at the origin. 
In the calculation, we set $M/T=10$,  and set (a) $\mu =0.5M$, (b) $\mu =M$, and (c) $\mu =1.5M$, respectively. 
Note $\varphi_{b,\vix}=-\varphi_{r_\vix}-\varphi_{g,\vix}$. 
Due to the P-H symmetry, the result in (c) is (almost) the same as that in (a) up to the total scale factor. 
}
\label{L_F_figure}
\end{figure}

In Fig.~\ref{L_F_figure}, the real part ${\rm Re}[{L}_{\rm F}]$ of EPLMWO is shown in $\varphi_{r,\vix}$--$\varphi_{g,\vix}$ plane. 
It should be noted that, in addition to points (configurations) shown in Fig.~\ref{L_F_figure}, there are the configurations that are obtained 
from configurations presented in Fig.~\ref{L_F_figure}
by performing the Weyl-symmetry transformation.   
It is seen that ${\rm Re}[{L}_{\rm F}]$ takes a minimum at the origin where $P_x=1$. 
Therefore, probabilistically, the deconfinement state is more favorable than the confinement state if only the fermion potential is considered. 
Note that $Z_3$-images of the origin, namely, points $(\varphi_{r,\vix},\varphi_{g,\vix})=(2\pi /3,2\pi /3), (-2\pi /3,-2\pi /3)$ are not favored by the fermion potential, since $Z_3$ symmetry is explicitly broken. 

The qualitative properties of $L_{\rm F}$, mentioned above, is independent of $\mu$, but  
the relative energy-difference ratio 
$R\equiv ({\rm Re}[L_{\rm F}({\rm max})-L_{\rm F}({\rm min})])/|{\rm Re}[L_{\rm F}({\rm min})]|$ between maximum and minimum values of ${\rm Re}[L_{\rm F}]$ is small when $\mu =M$.

%%%%%%%%%%%%%%%%%%%%%%
%%%%%%%%%%%%

\begin{figure}[htbp]
\begin{center}
\includegraphics[width=0.4\textwidth]{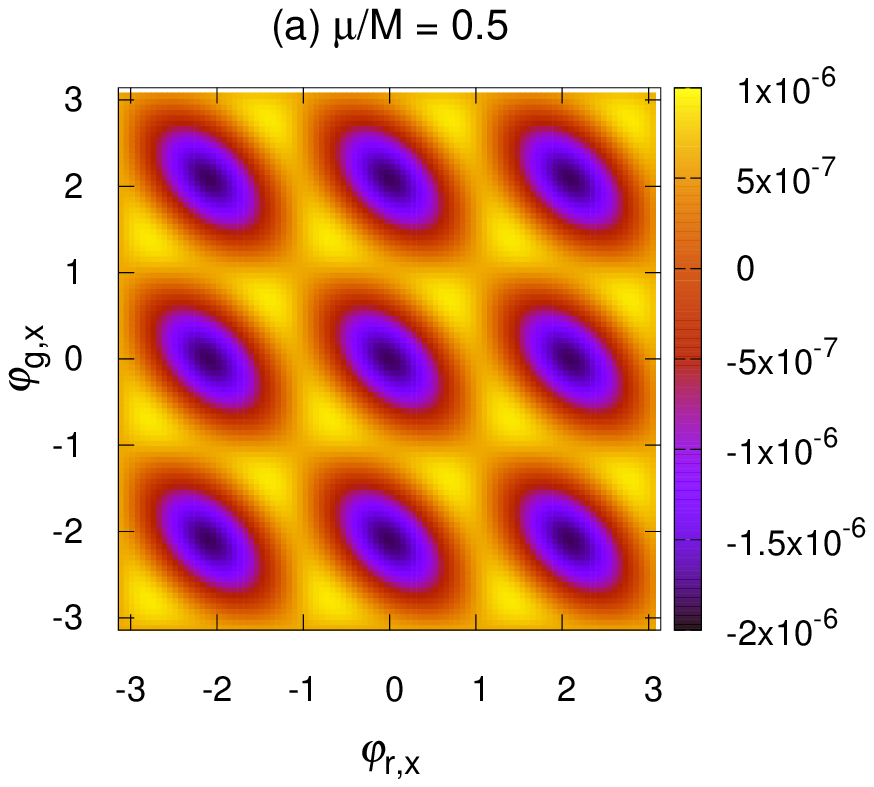}
\includegraphics[width=0.4\textwidth]{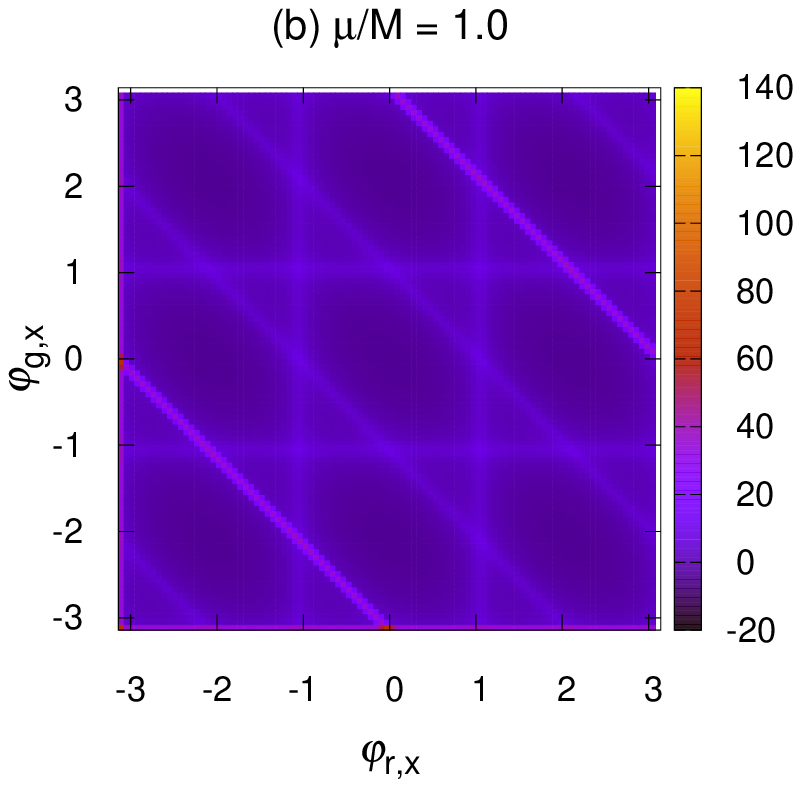}
\includegraphics[width=0.4\textwidth]{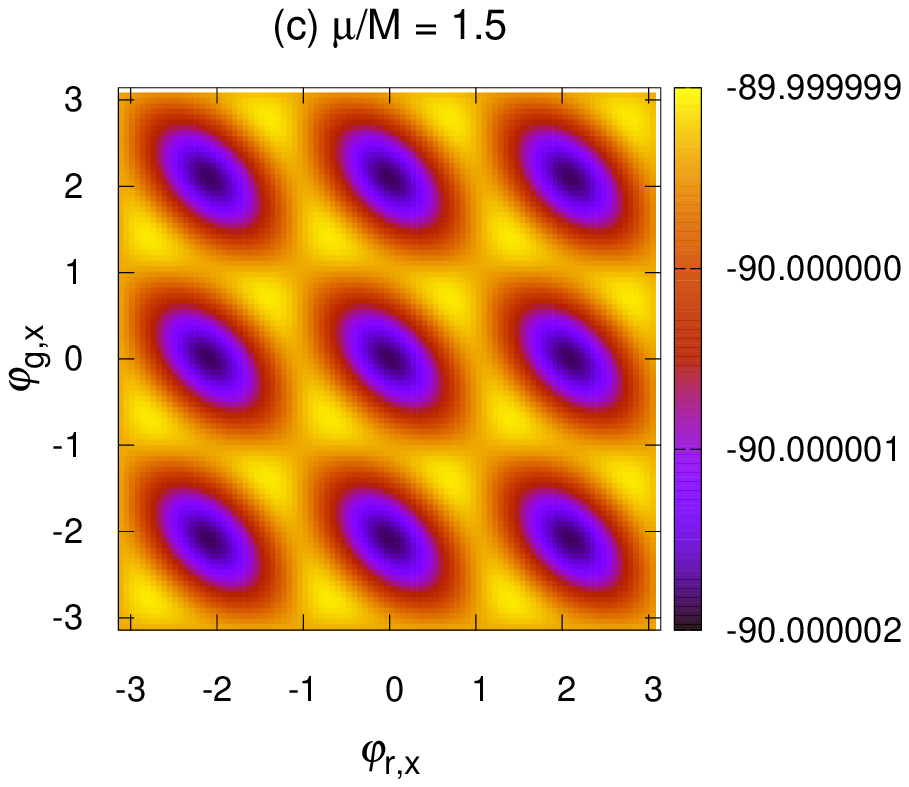}
\end{center}
 \caption{${\rm Re}[{L}_{{\rm F},Z_3}]$ 
in $\varphi_{r,\vix}$--$\varphi_{g,\vix}$ 
plane for the case of $Z_3$-EPLM.
 %when $Z_3$-EPLM is used. 
The fermion potential Largangian takes minimum at the confinement and deconfinement points. 
In the calculation, we set $M/T=10$,  and set (a) $\mu =0.5M$, (b) $\mu = M$, and (c) $\mu =1.5M$, respectively. 
Note $\varphi_{b,\vix}=-\varphi_{r_\vix}-\varphi_{g,\vix}$. 
Due to the P-H symmetry, the result in (c) is (almost) the same as that in (a) up to the total scale factor. 
}
\label{L_F_Z3_figure}
\end{figure}

Figure~\ref{L_F_Z3_figure} shows ${\rm Re}[{L}_{{\rm F},Z_3}]$ of $Z_3$-EPLM 
in $\varphi_{r,\vix}$--$\varphi_{g,\vix}$ plane. 
There are nine minimum points for each panel. 
The minimum points $(\varphi_{r,\vix},\varphi_{g,\vix})=(-2\pi /3,-2\pi /3),(0,0),(2\pi /3,2\pi /3)$ correspond to the deconfinement state 
and the other six points correspond to confinement state. 
These nine states are degenerate. 
The relative energy-difference ratio $R$ between maximum and minimum values of ${\rm Re}[L_{{\rm F},Z_3}]$ is very small when $\mu =M$.   
 
%%%%%%%%%%%%%%%%%%%%%%
%%%%%%%%%%%%}

\begin{figure}[htbp]
\begin{center}
\includegraphics[width=0.45\textwidth]{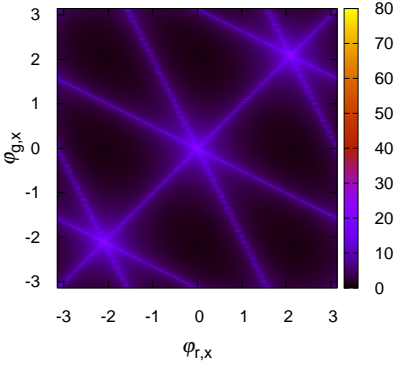}
\end{center}
 \caption{The effective potential $L_{\rm H}$ induced from the Haar measure is shown in $\varphi_{r,\vix}$--$\varphi_{g,\vix}$ plane. 
$L_{\rm H}$ takes minimum at the confinement points. 
}
\label{HaarMeasure}
\end{figure}

The degeneracy of the confinement and deconfinement states in the fermion effective action has a very important meaning. 
It means only the pure gauge contribution determines which configuration is probabilistically favorable. 
Figure~\ref{HaarMeasure} shows $L_{{\rm H}}$ in $\varphi_{r,\vix}$--$\varphi_{g,\vix}$ plane. 
 It is seen that $L_{{\rm H}}$ has a minimum at confinement points. 
Hence, the confinement state is favored in $Z_3$-EPLM with $N_s=1$, since there is no kinetic term in that case.

In the case of EPLMWO, even in the case of $N_s=1$, 
the situation is more complicated. 
In this case, $S_{\rm H}$ favors the confinement state, while $S_{\rm F}$ does the deconfinement one as mentioned above. 
Which state is favored? It depends on parameters taken in the model. 
If $M$ is large and $\mu$ is smaller than $M$, the confinement state is favorably realized, since the difference $\Delta L_{\rm F,R}\equiv {\rm Re}[L_{\rm F}(P_\vix=0)-L_{\rm F}(P_\vix =1)]$ is suppressed by the large $M$.  
Meanwhile, if $\mu$ is larger than $M$,  the contribution of $\Delta L_{\rm F,R}$ may be large enough to realize the deconfinement state.  

In EPLM with $N_s>1$, the gluon kinetic term $S_{\rm G}$ with parameter $\kappa$ exists. 
Since $S_{\rm G}$ favors an ordered deconfinement configuration, the 
deconfinement phase is realized as an ordered phase when $\kappa$ is large.  

It should be remarked that, when $N_s>1$, the spatial average value $\bar{P}={1\over{V}}\displaystyle{\sum_{\vix}}P_\vix$ can be zero even if $P_\vix \neq 0$. 
In fact, as seen later in Sec.~\ref{Result_mu=0}, such a cancellation does happen at small $\kappa$ and $\mu$. 
In this case, the confinement phase appears as a random phase in which $P_\vix$ fluctuates largely. 
Note that, in EPLM, the confinement state $P_\vix =0$ and the deconfinement state $P_\vix =1$ preserve the reality of the action. Particularly for $Z_3$-EPLM, 
the $Z_3$-images of $P_\vix =1$ also preserve the reality. 
However, fluctuations of $P_\vix$ in the random phase cause 
the sign problem when $\mu$ increases.

%%%%%%%%%%%%%%%%%%%%%%
%%%%%%%%%%%%

%\newpage

%\end{document}

%%%%%%%%%%%%%%%%%%%%%%%%%%%%%%%%%%%%%%%%%%%%%%%%%%%%%%%%%%%%%%%%
\section{Reweighting methods}
\label{Section_RW}
%%%%%%%%%%%%%%%%%%%%%%%%%%%%%%%%%%%%%%%%%%%%%%%%%%%%%%%%%%%%%%%%
\subsection{Phase quenched approximation and reweighting method}
\label{subs_PQA}
%%%%%%%%%%%%%%%%%%%%%%%%%%%%%%%%%%%%%%%%%%%%%%%%%%%%%%%%%%%%%%%%%

Since, at finite $\mu$, both EPLMWO and $Z_3$-EPLM have the sign problem in its path integral formalism, we use the phase quenched approximation (PQA).  
Using the approximate probability function $F^\prime /Z^\prime$, we 
calculate the approximate average value of a physical quantity $O$ as 
%%%%%%%%%%%%%%%%
\begin{eqnarray}
&&\langle O \rangle^\prime 
={\int {\cal D}\varphi_rD\varphi_g O(\varphi_{r,\vix},\varphi_{g,\vix} )F^\prime (\varphi_{r,\vix},\varphi_{g,\vix})\over{Z^\prime }};
\nonumber\\
&&F^\prime (\varphi_{r,\vix},\varphi_{g,\vix} )= e^{-S_{\rm 0}-S_{\rm F,R}}, 
\nonumber\\
&&Z^\prime =\int {\cal D}\varphi_rD\varphi_g F^\prime (\varphi_{r,\vix},\varphi_{g,\vix} ), 
\label{EPLM_phase_quench}
\end{eqnarray}
%%%%%%%%%%%%%%
where $S_{\rm 0}=S_{\rm G}+S_{\rm H}$ and $S_{\rm F,R}={\rm Re}[S_{\rm F}]$. 
The phase factor $W^\prime$ is given by
%%%%%%%%%%%%%%%%
\begin{eqnarray}
W^\prime &=&{Z\over{Z^\prime}}
={\int {\cal D}\varphi_r{\cal D}\varphi_g \left[e^{-iS_{\rm F,I}}\right]F^\prime (\varphi_{r,\vix},\varphi_{g,\vix})\over{Z^\prime }}
\nonumber\\
&=&\langle e^{-iS_{\rm F,I}} \rangle^\prime
=\left\langle {F\over{F^\prime}} \right\rangle^\prime; 
\nonumber\\
\nonumber\\
&&F (\varphi_{r,\vix},\varphi_{g,\vix} )= e^{-S_{\rm 0}-S_{\rm F}}=e^{-S_{\rm 0}-S_{\rm F,R}-iS_{\rm F,I}}, 
\nonumber\\
&&Z =\int {\cal D}\varphi_rD\varphi_g F (\varphi_{r,\vix},\varphi_{g,\vix} ),  
\label{phase_factor}
\end{eqnarray}
%%%%%%%%%%%%%%%%
where $S_{\rm F,I}={\rm Im}[S_{\rm F}]$. 
Taking the partial average of two configurations $(\varphi_{r,\vix},\varphi_{g,\vix})$ and $(\varphi_{r,\vix}^\prime,\varphi_{g,\vix}^\prime)$ that satisfy $(\varphi_{r,\vix}^\prime,\varphi_{g,\vix}^\prime)=(-\varphi_{r,\vix},-\varphi_{g,\vix})$, we obtain~\cite{Hirakida} 
%%%%%%%%%%%%%%%%
\begin{eqnarray}
W^\prime
&=&{\int {\cal D}\varphi_r{\cal D}\varphi_g\left[\cos{(S_{\rm F,I})}\right]F^\prime (\varphi_{r,\vix},\varphi_{g,\vix})\over{Z^\prime }}
\nonumber\\
&=&\langle \cos{(S_{\rm F,I})} \rangle^\prime, 
\label{phase_factor_2}
\end{eqnarray}
%%%%%%%%%%%%%%%%
since $S_{\rm F,I}(\varphi_{r,\vix}^\prime,\varphi_{g,\vix}^\prime)
=-S_{\rm F,I}(\varphi_{r,\vix},\varphi_{g,\vix})$. 
Hence the phase factor is real and $|W^{\prime} | \leq 1$ is satisfied.  
Using the reweighting method, one can obtain the true expectation value $\langle O \rangle$ as 
%%%%%%%%%%%%%%
\begin{eqnarray}
&&\langle O \rangle 
={\int {\cal D}\varphi_r{\cal D}\varphi_g \left[ O(\varphi_{r,\vix},\varphi_{g,\vix})\right]F(\varphi_{r,\vix},\varphi_{g,\vix})\over{Z}}
\nonumber\\
&&={\int {\cal D}\varphi_r{\cal D}\varphi_g \left[ Oe^{-iS_{{\rm F,I}}}\right]F^{\prime}\over{Z^{\prime}}}\times {Z^{\prime} \over{Z}}
\nonumber\\
&&={\langle Oe^{-iS_{{\rm F,I}}}\rangle^\prime\over{W^{\prime}}}. 
\label{rw_phase_PQA}
\end{eqnarray}
%%%%%%%%%%%%%%
In this paper, we refer this reweighting method as ``phase quenched reweighting (PQRW)''. 

If $\cos{(S_{\rm F,I})}$ can have a negative sign, the absolute value of the 
phase factor (\ref{phase_factor_2}) becomes small due to the cancellation between the configurations with positive $\cos{(S_{\rm F,I})}$ and those with negative $\cos{(S_{\rm F,I})}$. 
In actual simulations, the smallness of the absolute value of $W^\prime$ causes large errors in the division (\ref{rw_phase_PQA}) and makes the calculation results unreliable. 
Hence, in PQRW, the phase factor $W^\prime$ indicates how serious the sign problem is.  

%%%%%%%%%%%%%%%%%%%%%%%%%%%%%%%%%%%%%%%%%%%%%%%%
\subsection{Sign problem in EPLM}
\label{EPLM0_sign}
%%%%%%%%%%%%%%%%%%%%%%%%%%%%%%%%%%%%%%%%%%%%%%%%

\begin{figure}[htbp]
\begin{center}
\includegraphics[width=0.45\textwidth]{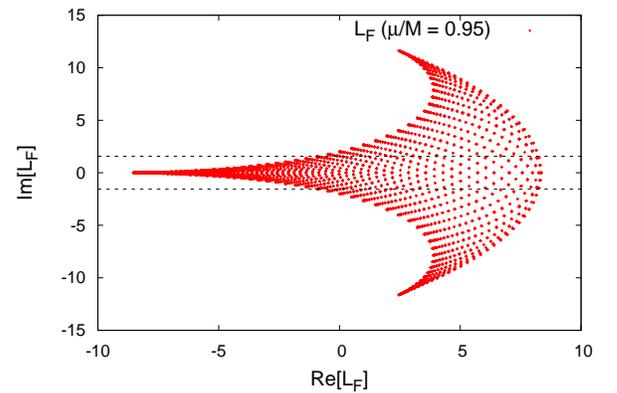}
\end{center}
\caption{The 
${\rm Re}[{L}_{\rm F}]$-${\rm Im}[{L}_{\rm F}]$ 
relation  
in the 3 flavor EPLMWO. 
We set $M/T=10$ and $\mu /M=0.95$. }
\label{LR-LI_nf3}
\end{figure}

\begin{figure}[htbp]
\begin{center}
\includegraphics[width=0.45\textwidth]{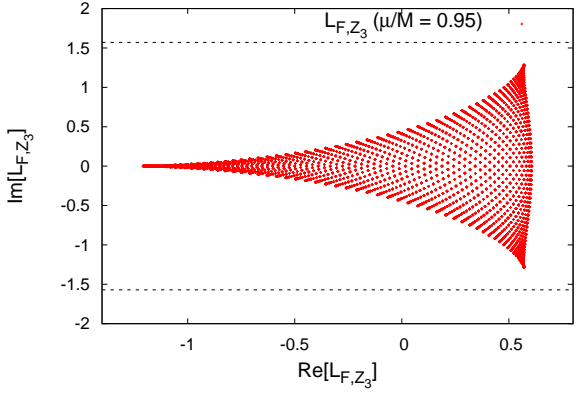}
\end{center}
\caption{ The 
${\rm Re}[{L}_{{\rm F},Z_3}]$-${\rm Im}[{L}_{{\rm F},Z_3}]$ 
 relation   
in $Z_3$-EPLM.  
We set $M/T=10$ and $\mu /M=0.95$. 
}
\label{LR-LI_z3}
\end{figure}

In order  to see the fact that the sign problem is milder in $Z_3$-EPLM than in EPLMWO, we examine the imaginary part of the fermionic potential. 
Figures~\ref{LR-LI_nf3} and~\ref{LR-LI_z3} show the ${\rm Re}[{L}_{\rm F}]$-${\rm Im}[{L}_{\rm F}]$ relation 
for three-flavor EPLMWO and $Z_3$-EPLM, respectively. 
We set $\mu =0.95M$, since the sign problem is severest in the vicinity of $\mu =0.95M$ (or $1.05M$), as seen in the next section. 
Both the figures have qualitatively the same structure, but their physical meanings are much different. 
In $Z_3$-EPLM, the confinement state $P_\vix =0$ and the deconfinement states $P_\vix =1,e^{\pm i2\pi /3}$ are degenerate on the left vertex where the sign problem is absent and $e^{-{\rm Re}[L_{{\rm F},Z_3}]}$ is largest. Meanwhile, 
in EPLMWO only the state $P_\vix =1$ is present 
on the left vertex. 
More generally, the states $P_\vix =b$ and $P\vix=be^{\pm i2\pi/3}~~~(b=-1/3\sim 1)$  are degenerate in $Z_3$-EPLM, but not in EPLMWO. 
There is a tendency that, in both of EPLMWO and $Z_3$-EPLM, the absolute value of 
${\rm Im}[{L}_{\rm F}]$ becomes large, when the absolute value of 
${\rm Re}[{L}_{\rm F}]$ is large. 
The absolute values of ${\rm Re}[{L}_{\rm F}]$ 
and ${\rm Im}[{L}_{\rm F}]$ themselves, however,  
are much smaller in $Z_3$-EPLM than in EPLMWO.  
The latter property comes from the fact that  the mass $M$ present 
in $L_{{\rm F},Z_3}$ is always multiplied by a factor of 3  and 
the absolute value $|L_{{\rm F},Z_3}|$ is more suppressed than 
$|L_{\rm F}|$. 

%%%%%%%%%%%%%%%%%%%%%%
%%%%%%%%%%%%

%%%%%%%%%%%%%%%%%%%%%%
%%%%%%%%%%%%

In Fig. \ref{LR-LI_nf3} (Fig. \ref{LR-LI_z3}), we see that the maximum value of the absolute value of ${\rm Im}[{L}_{\rm F}]$ (${\rm Im}[{L}_{{\rm F},Z_3}]$) is 
larger (smaller) than ${\pi\over{2}}$ at $\mu/M=0.95$.  
This means that, at $\mu /M=0.95$, the sign of 
 $\cos{({\rm Im}[{L}_{{\rm F},Z_3}])}$ is always positive in $Z_3$-EPLM, while the sign of $\cos{({\rm Im}[{L}_{{\rm F}}])}$is not definite in EPLMWO.  
Hence, in $Z_3$-EPLM with $N_s=1$,  
there is no sign problem at $\mu /M=0.95$ when $N_s=1$.   
However, even in $Z_3$-EPLM, the absolute value of $S_{{\rm F},{\rm I}}=\sum_{\vix}{\rm Im}[L_{{\rm F},Z_3}({\bf x})]$ can be large and $\cos{({S}_{{\rm F},Z_3,{\rm I}})}$ can have a negative sign when $N_s$ increases. 
This causes the sign problem in $Z_3$-EPLM with larger $N_s$, although it is milder than in EPLMWO.

%%%%%%%%%%%%%%%%%%%%%%%%%%%%%%%%%%%%%%%%%%%%%%%%%%%%%%%%%
\subsection{Improved reweighting method}
\label{subs_IRM}
%%%%%%%%%%%%%%%%%%%%%%%%%%%%%%%%%%%%%%%%%%%%%%%%%%%%%%%%%

In order to improve PQRW, we assume that the realization probability of the configuration with $S_{\rm R}$ and $S_{\rm I}$ is well approximated by the probability distribution function proportional to $e^{-S_{\rm R}}e^{-\alpha S_{\rm I}^2}$,  where $\alpha$ is an appropriate parameter that may depend on $\kappa$, $\mu$ and $N_s$.
In fact, Ejiri studied  the distribution of the phase of the quark determinant  by using the Taylor expansion method 
and found that it can be well approximated by a Gaussian function~\cite{Ejiri_theta}. 
Similar analysis was made with the strong coupling expansion method~\cite{Ohnishi_preW}.   

We consider the following approximation: 
%%%%%%%%%%%%%%%%
\begin{eqnarray}
&&\langle O \rangle^{\prime\prime} 
={\int {\cal D}\varphi_r{\cal D}\varphi_g O(\varphi_{r,\vix},\varphi_{g,\vix})F^{\prime\prime} (\varphi_{r,\vix},\varphi_{g,\vix})\over{Z^{\prime\prime} }};
\nonumber\\
&&F^{\prime\prime} (\varphi_{r,\vix},\varphi_{g,\vix})= e^{-S_{\rm 0}-S_{\rm F,R}}e^{-\alpha S_{\rm F,I}^2},   
\nonumber\\
&&Z^{\prime\prime} =\int{\cal D}\varphi_r{\cal D}\varphi_g F^{\prime\prime} (\varphi_{r,\vix},\varphi_{g,\vix}), 
\label{IPQA}
\end{eqnarray}
%%%%%%%%%%%%%%
In actual simulations, we treated $\alpha$ as a variable parameter 
and searched a best value for $\alpha$. 

As shown in the case of PQRW, 
the ratio $W^{\prime\prime}=Z/Z^{\prime\prime}$ is given by
%%%%%%%%%%%%%%%%
\begin{eqnarray}
W^{\prime\prime} &=&{Z\over{Z^{\prime\prime}}}
={\int {\cal D}\varphi_r{\cal D}\varphi_g \left[e^{-iS_{\rm F,I}}e^{\alpha S_{\rm F,I}^2} \right]F^\prime (\varphi_{r,\vix},\varphi_{g,\vix})\over{Z^{\prime\prime} }}
\nonumber\\
&=&\langle e^{-iS_{\rm F,I}} e^{\alpha S_{\rm F,I}^2}\rangle^{\prime\prime}
=\left\langle {F\over{F^{\prime\prime}}} \right\rangle^{\prime\prime}.  
\label{phase_factor_improved}
\end{eqnarray}
%%%%%%%%%%%%%%%%
Taking the partial average of two configurations $(\varphi_{r,\vix},\varphi_{g,\vix})$ and $(\varphi_{r,\vix}^\prime,\varphi_{g,\vix}^\prime)$ that satisfy $(\varphi_{r,\vix}^\prime,\varphi_{g,\vix}^\prime)=(-\varphi_{r,\vix},-\varphi_{g,\vix})$, we obtain~\cite{Hirakida} 
%%%%%%%%%%%%%%%%
\begin{eqnarray}
W^{\prime\prime}
&=&{\int {\cal D}\varphi_r{\cal D}\varphi_g\left[\cos{(S_{\rm F,I})e^{\alpha S_{\rm F,I}^2}}\right]F^{\prime\prime} (\varphi_{r,\vix},\varphi_{g,\vix})\over{Z^{\prime\prime} }}
\nonumber\\
&=&\langle \cos{(S_{\rm F,I})} e^{\alpha S_{\rm F,I}^2}\rangle^{\prime\prime}, 
\label{phase_factor_improved_2}
\end{eqnarray}
%%%%%%%%%%%%%%%%
since $S_{\rm F,I}(\varphi_{r,\vix}^\prime,\varphi_{g,\vix}^\prime)
=-S_{\rm F,I}(\varphi_{r,\vix},\varphi_{g,\vix})$. 
Unlike the case of PQRW, 
the condition $|W^{\prime\prime} | \leq 1$ 
is not ensured in this case. 
If $F^{\prime\prime}$ is a good approximate probability distribution function,  $W^{\prime\prime} \sim \tilde{W}^{\prime\prime}\equiv \langle e^{\alpha S_{\rm I}^2}\rangle^{\prime\prime}\ge 1$ is expected instead of $W^{\prime\prime} \sim 1$. 
Using the reweighting method, we can obtain the true expectation value $\langle O \rangle$ as 
%%%%%%%%%%%%%%
\begin{eqnarray}
&&\langle O \rangle 
={\int {\cal D}\varphi_r{\cal D}\varphi_g \left[ O(\varphi_{r,\vix},\varphi_{g,\vix})\right]F(\varphi_{r,\vix},\varphi_{g,\vix})\over{Z}}
\nonumber\\
&&={\int {\cal D}\varphi_r{\cal D}\varphi_g \left[ O e^{-iS_{{\rm F,I}}}e^{\alpha S_{\rm F,I}^2}\right]F^{\prime\prime} \over{Z^{\prime\prime}}}\times {Z^{\prime\prime} \over{Z}}
\nonumber\\
&&={\langle O e^{-iS_{{\rm F,I}}}e^{\alpha S_{\rm F,I}^2}\rangle^{\prime\prime}\over{W^{\prime\prime}}}. 
\label{rw_phase_IPQA}
\end{eqnarray}
%%%%%%%%%%%%%%
This reweighting method is referred to as 
``improved phase quenched reweighting 
(IPQRW)'' in this paper.

%%%%%%%%%%%%%%%%%%%%%%%%%
\subsection{Observables}
%%%%%%%%%%%%%%%%%%%%%%%%%

In numerical calculations, we consider the following quantities as observables. 
First, we consider the spatial average $\bar{P}$ of $P_\vix$: 
Namely,   
%%%%%%%%%%%%%%%%%%%%%%%%%%%%%
\begin{eqnarray}
\bar{P} ={1\over{V}}\sum_\vix P_\vix. 
\label{Phi_x_average}
\end{eqnarray}
%%%%%%%%%%%%%%%%%%%%%%%%%%%%%
We calculate the expectation value of $|\bar{P} |$, since this quantity defines random and ordered states in the system and plays a role of the expectation value of Polyakov-line in QCD. 

Another quantity is the quark number density given by 
%%%%%%%%%%%%%%%%%%%%%%%%%%%%%
\begin{eqnarray}
n_q &=&{1\over{\beta V}}{\partial (\log{Z})\over{\partial \mu}}
={1\over{V}}{\partial \log{Z}\over{\partial {\hat{\mu}}}},  
\label{n_density}
\end{eqnarray}
%%%%%%%%%%%%%%%%%%%%%%%%%%%%%
where $\hat{\mu}=\mu /T$. 
LQCD at finite $\mu$ has the problem on the early onset of quark number density~\cite{Barbour} or the baryon Silver Blaze problem~\cite{Cohen}. 
The quark number density $n_q$ should be zero at $T=0$ for $\mu<M_{\rm N}/3$, 
where $M_{\rm N}$ is the nucleon mass.  
However, it becomes finite in LQCD calculations for $\mu >m_\pi/2$ when PQRW is used, where $m_\pi$ is the pion mass. 
Therefore, $n_q$ is also useful to check whether our simulations are 
reliable or not.

%%%%%%%%%%%%%%%%%%%%%%%%%%%%%%%%%%%%%%%%%%%%%%%%%%%%%%%%
%%%% numerical results
%%%%%%%%%%%%%%%%%%%%%%%%%%%%%%%%%%%%%%%%%%%%%%%%%%%%%%%%
\section{Numerical results}
\label{numerical}
%%%%%%%%%%%%%%%%%%%%%%%%%%%%%%%%%%%%%%%%%%%%%%%%%%%%%%%%

In this section, we present numerical evaluations of the phase factor $W^\prime$ (or the factor $W^{
\prime\prime}$) and the expectation values of $|\bar{P}|$ and $n_q$. 
Our simulations were made by using the standard Monte Carlo algorithm. 
We consider three cases of $N_s=6$, 8, 12. 
We also set $M/T=10$ in this section, unless otherwise mentioned.    

%%%%%%%%%%%%%%%%%%%%%%%%%%%%%%%%%%%%%%%%%%%%%%%%%
\subsection{Results at $\mu =0$}
\label{Result_mu=0}
%%%%%%%%%%%%%%%%%%%%%%%%%%%%%%%%%%%%%%%%%%%%%%%%%

\begin{figure}[htbp]
\begin{center}
\includegraphics[width=0.35\textwidth]{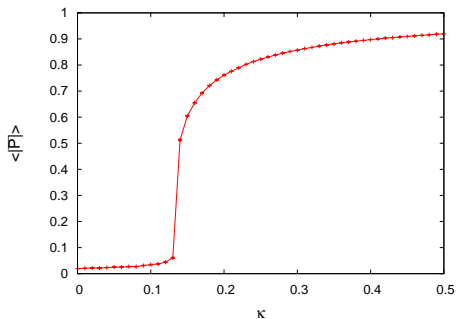}
\end{center}
\caption{The $\kappa$-dependence of expectation value $\langle |\bar{P} | \rangle$ at $\mu =0$ in $Z_3$-EPLM. 
We set $M/T=10$, $\mu =0$ and $N_s=6$. 
}
\label{Fig_kappa_P_mu0_Z3EPLM}
\end{figure}

First we discuss the case of $\mu =0$, since LQCD simulations have 
no sign problem there. 
Hence, the reweighting procedure is not needed. 
In Fig.~\ref{Fig_kappa_P_mu0_Z3EPLM}, the $\kappa$-dependence of $\langle |\bar{P}| \rangle$  is shown for $Z_3$-EPLM.    
As $\kappa$ becomes large, $\langle |\bar{P}| \rangle$ increases. 
There is a rapid change of $\langle |\bar{P}| \rangle$ at $\kappa \sim 0.13$. 
The change seems to 
show a first-order phase transition, 
but we postpone our conclusion on the order 
since $N_s$ in our simulations is not large 
enough to determine the order of the transition. 

\begin{figure}[htbp]
\begin{center}
\includegraphics[width=0.4\textwidth]{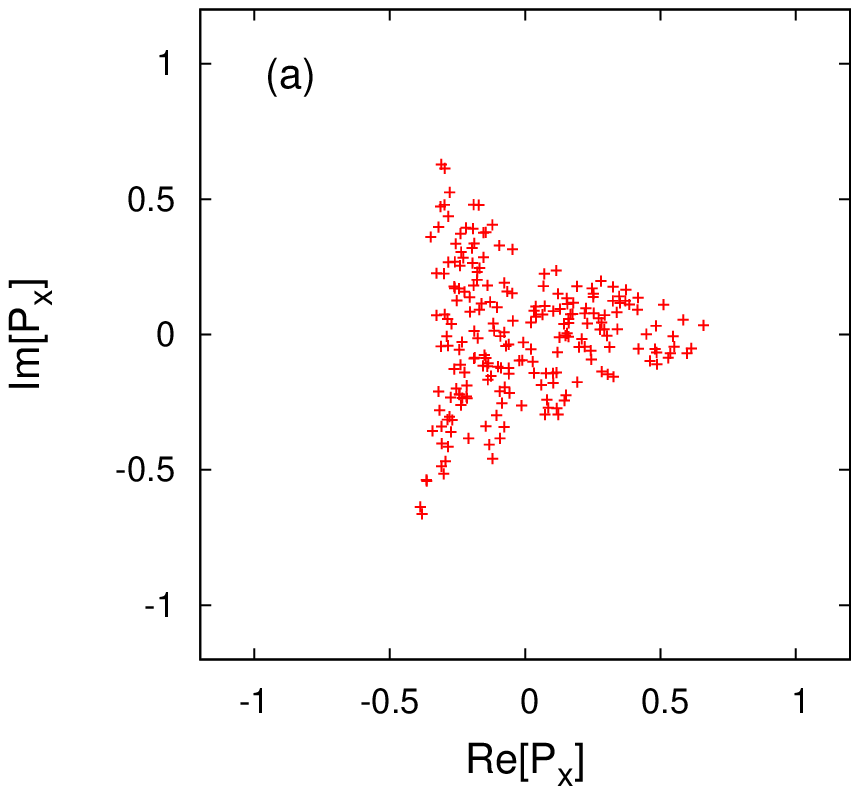}
\includegraphics[width=0.4\textwidth]{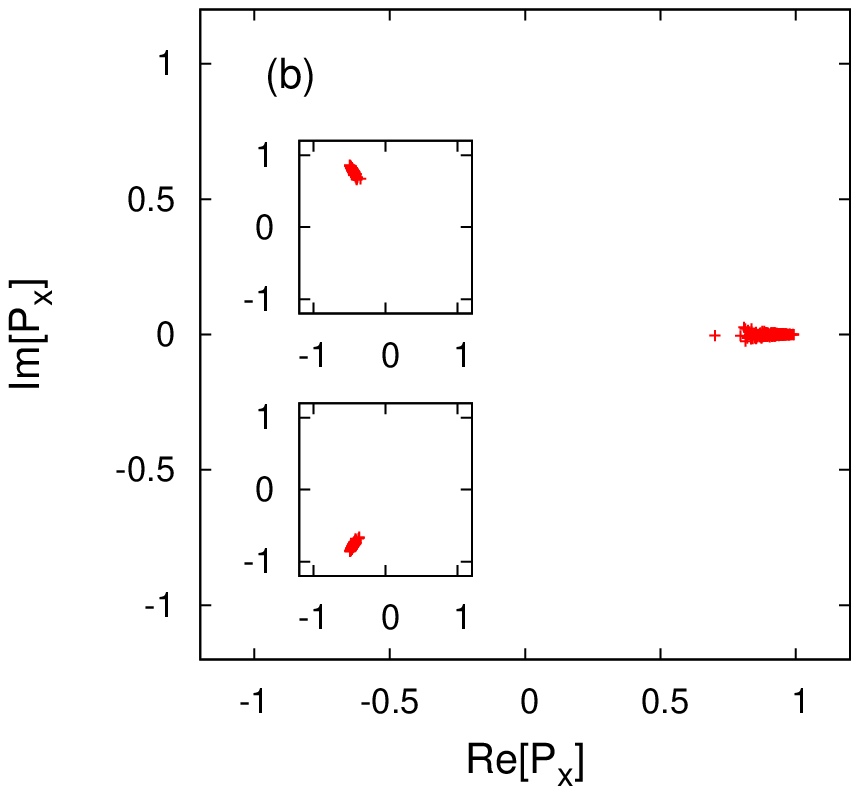}
\end{center}
\caption{The scatter plot of $P_\vix$ at (a) $\kappa=0$ and (b) $\kappa=0.5$ in $Z_3$-EPLM. 
We set $M/T=10$, $\mu=0$ and $N_s=6$.  
Note that $P_\vix$ in one configuration is plotted.  
In this configuration, $P_\vix$ is concentrated in the vicinity of $P_\vix =1$ when $\kappa=0.5$.  
In the insets, configurations of $P_\vix$ concentrated in the vicinity of $P_\vix =e^{i2\pi /3}$ and $P_\vix =e^{-i2\pi /3}$ are shown. 
}
\label{P-scatter}
\end{figure}

Figure~\ref{P-scatter} shows a sample of scatter plots of $P_\vix$ at $\mu =0$ 
for the case of $Z_3$-EPLM. 
Note that any configuration sum is not taken in Fig.~\ref{P-scatter}. 
Even in one configuration, at $\kappa =0$, $P_\vix$ widely distributes in the complex plane and $P_\vix =0$ is not always ensured. 
However, the spatial average $\bar{P}$ almost vanishes due to the cancellation among variables $P_\vix$ on different lattice sites.   
At large $\kappa$, an ordered configuration of $P_\vix$ is favored, and $\bar{P}_\vix \approx 1, e^{\pm i 2\pi /3}$ is realized.

%%%%%%%%%%%%%%%%%%%%%%
%%%%%%%%%%%%%%%%%%%%%%%%%%%%%%%%%%

%%%%%%%%%%%%%%%%%%%%%%
%%%%%%%%%%%%%%%%%%%%%%%%%%%%%%%%%%

Similar result is obtained in EPLMWO, since 
the effect of the fermion action is small at $\mu =0$ and $Z_3$ symmetry is almost preserved when $M$ is large. 
If we use smaller mass $M=5T$, the configurations of $P_\vix$ at large $\kappa$ are concentrated only in the vicinity of $P_\vix =1$, since $Z_3$ symmetry is largely broken in EPLMWO with smaller $M$.

%%%%%%%%%%%%%%%%%%%%%%%%%%%%%%%%%%%%%%%%%%%%%%%%%
\subsection{Phase quenched reweighting}
%%%%%%%%%%%%%%%%%%%%%%%%%%%%%%%%%%%%%%%%%%%%%%%%%

EPLMWO and $Z_3$-EPLM have the sign problem for finite $\mu$. 
Hence, we use the reweighting method here.

\begin{figure}[htbp]
\begin{center}
\includegraphics[width=0.4\textwidth]{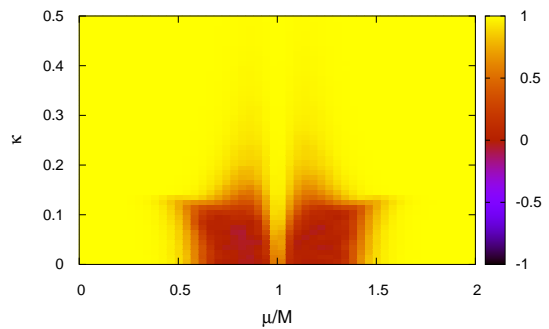}
\end{center}
\caption{The phase factor $W^\prime$ in EPLMWO, where PQRW is used. 
We set $N_s=6$. 
}
\label{Fig_Phase_factor_ordinary}
\end{figure}

Figure~\ref{Fig_Phase_factor_ordinary} shows the phase factor for EPLMWO, where PQRW is used. 
Due to the P-H symmetry, in $\mu$--$\kappa$ plane, the result is almost symmetric with respect to the line $\mu =M$. 
It is seen that the sign problem is serious when $\mu /M=0.5\sim 1.5$ and $\kappa <0.15$. 
For $\mu =M$, however, $S_{\rm F}$ is real and the phase factor is 1, 
as already mentioned in Sec. \ref{EPLM_A}; note that these properties are 
not clearly seen  in Fig.~\ref{Fig_Phase_factor_ordinary}.  

%%%%%%%%%%%%%%%%%%%%%%
%%%%%%%%%%%%

\begin{figure}[htbp]
\begin{center}
\includegraphics[width=0.4\textwidth]{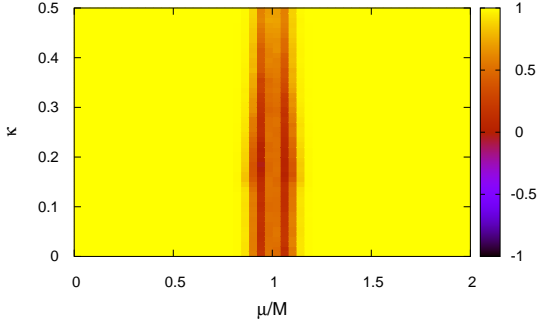}
\end{center}
\caption{The phase factor $W^\prime$ in $Z_3$-EPLM, where PQRW is used. 
We set $N_s=6$. 
}
\label{Fig_Phase_factor_Z3}
\end{figure}

Figure~\ref{Fig_Phase_factor_Z3} shows the phase factor for $Z_3$-EPLM, 
where PQRW is used. 
The phase factor is small only in the region $\mu /M=0.85\sim 1.15$. 
(As in the case of EPLMWO, $W^\prime =1$ just on the line $\mu =M$. )   
For $\mu /M=0.5\sim 0.85$ ($1.15\sim 1.5$) and $\kappa <0.12$, the sign problem is considerably milder in $Z_3$-EPLM than in EPLMWO. 
On the contrary, when $\mu /M=0.85\sim 1.15$ and $\kappa >0.12$, the phase factor is somewhat smaller in $Z_3$-EPLM than in EPLMWO. 
This may be originated in the fact that, as already seen in Sec.~\ref{EPLM0}, $L_{{\rm F},Z_3}$ is almost flat near $\mu =M$ and $P_x$ can fluctuate considerably.

%%%%%%%%%%%%%%%%%%%%%%
%%%%%%%%%%%%

%%%%%%%%%%%%%%%%%%%%%%%%%%%%%%%%%%%%%%%%%%%%%%%%%
\subsection{Improved phase quenched reweighting}
%%%%%%%%%%%%%%%%%%%%%%%%%%%%%%%%%%%%%%%%%%%%%%%%%

%%%%%%%%%%%%%%%%%%%%%%
\begin{figure}[htbp]
\begin{center}
\includegraphics[width=0.4\textwidth]{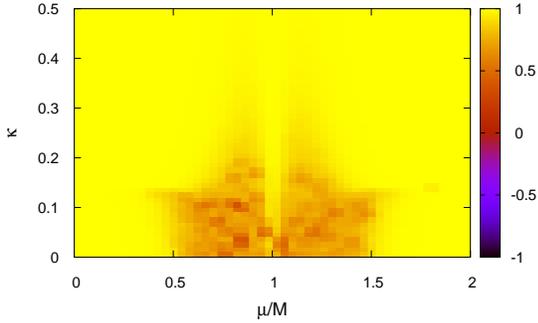}
\end{center}
\caption{The ratio $W^{\prime\prime} /\tilde{W}^{\prime\prime}$ in EPLMWO, where IPQRW is used. 
We set $\alpha =3.5$ and $N_s=6$. 
}
\label{pf_EPLM_I}
\end{figure}
%%%%%%%%%%%%

%%%%%%%%%%%%%%%%%%%%%%
\begin{figure}[htbp]
\begin{center}
\includegraphics[width=0.4\textwidth]{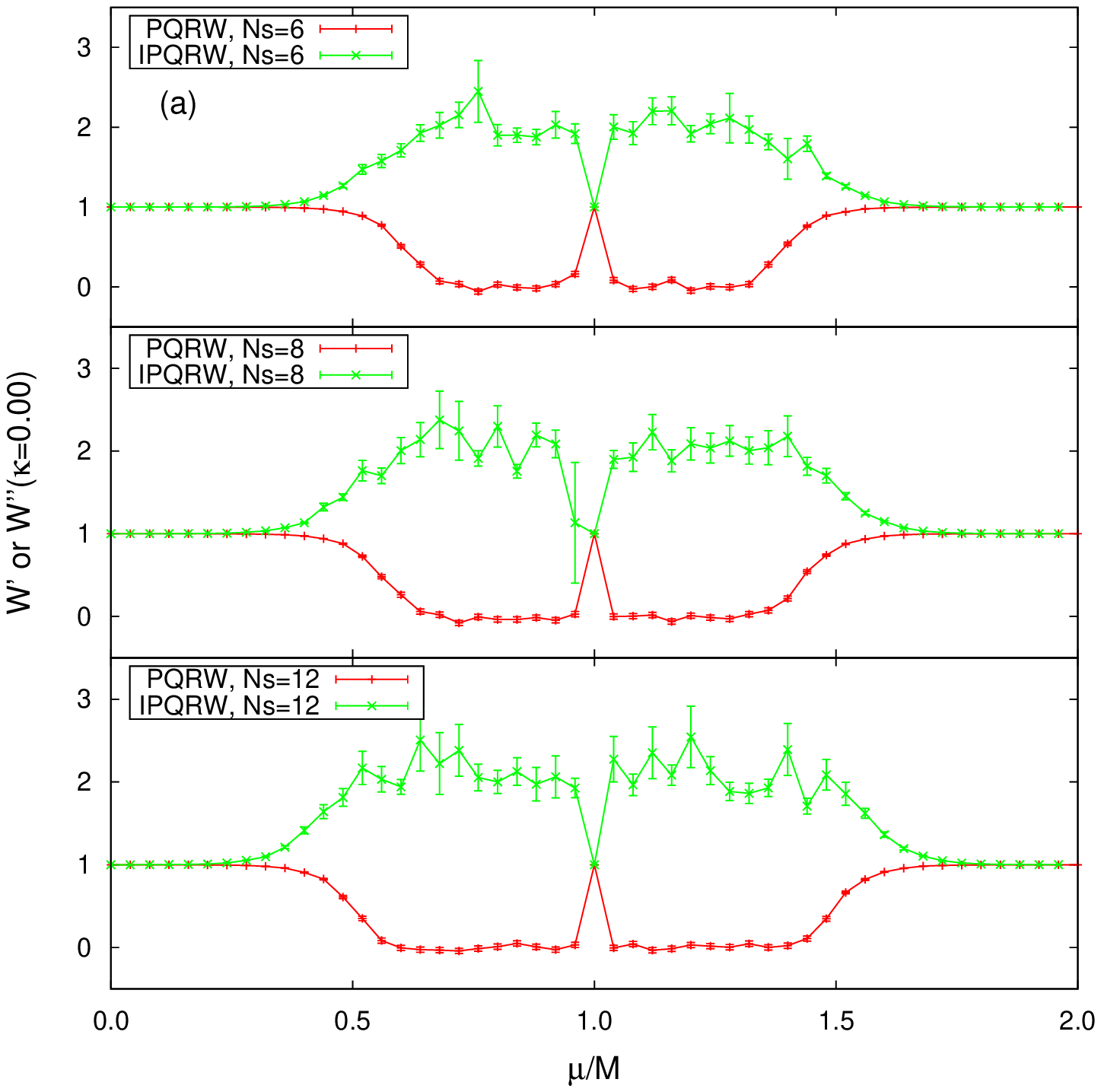}
\includegraphics[width=0.4\textwidth]{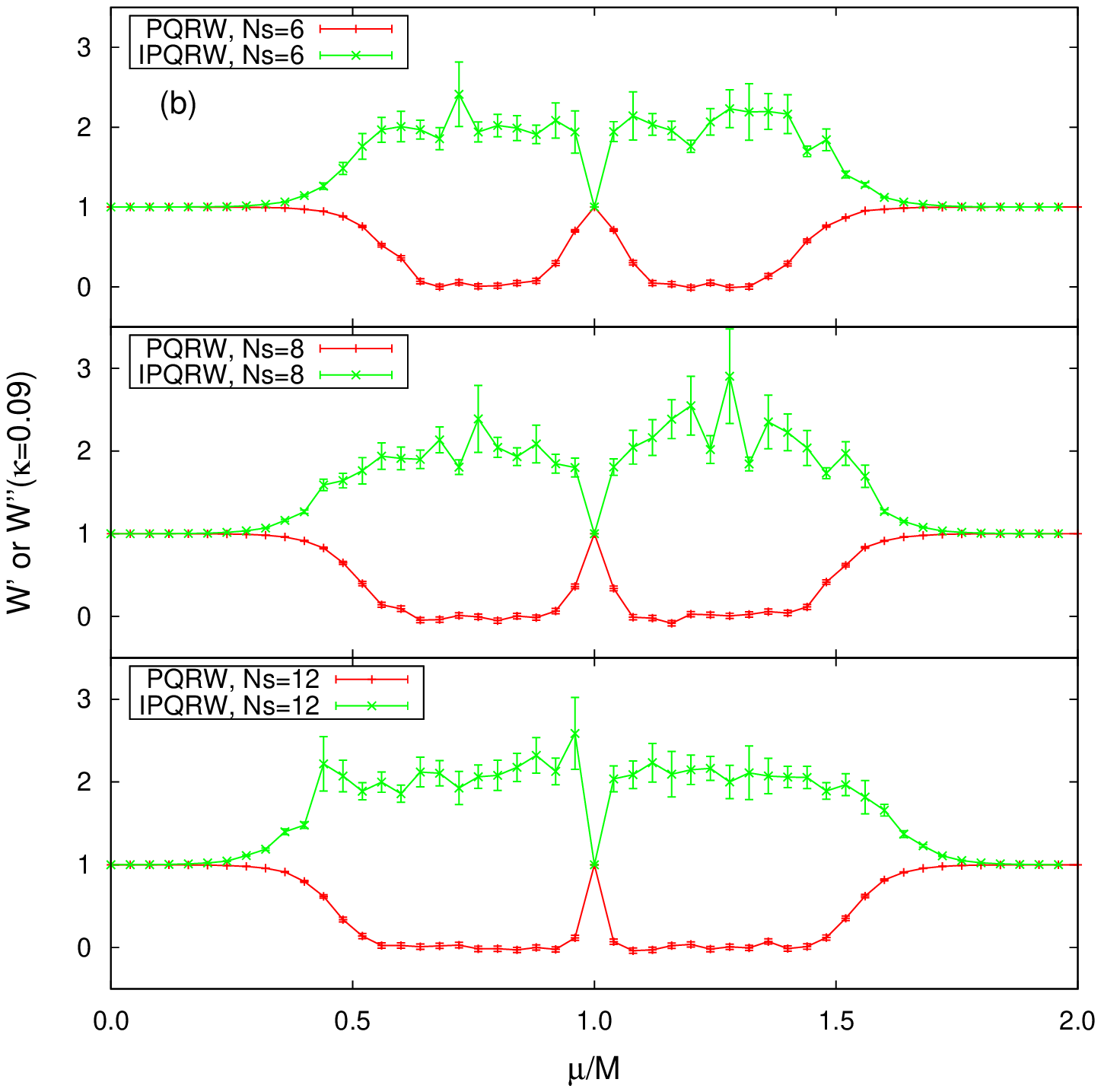}
\includegraphics[width=0.4\textwidth]{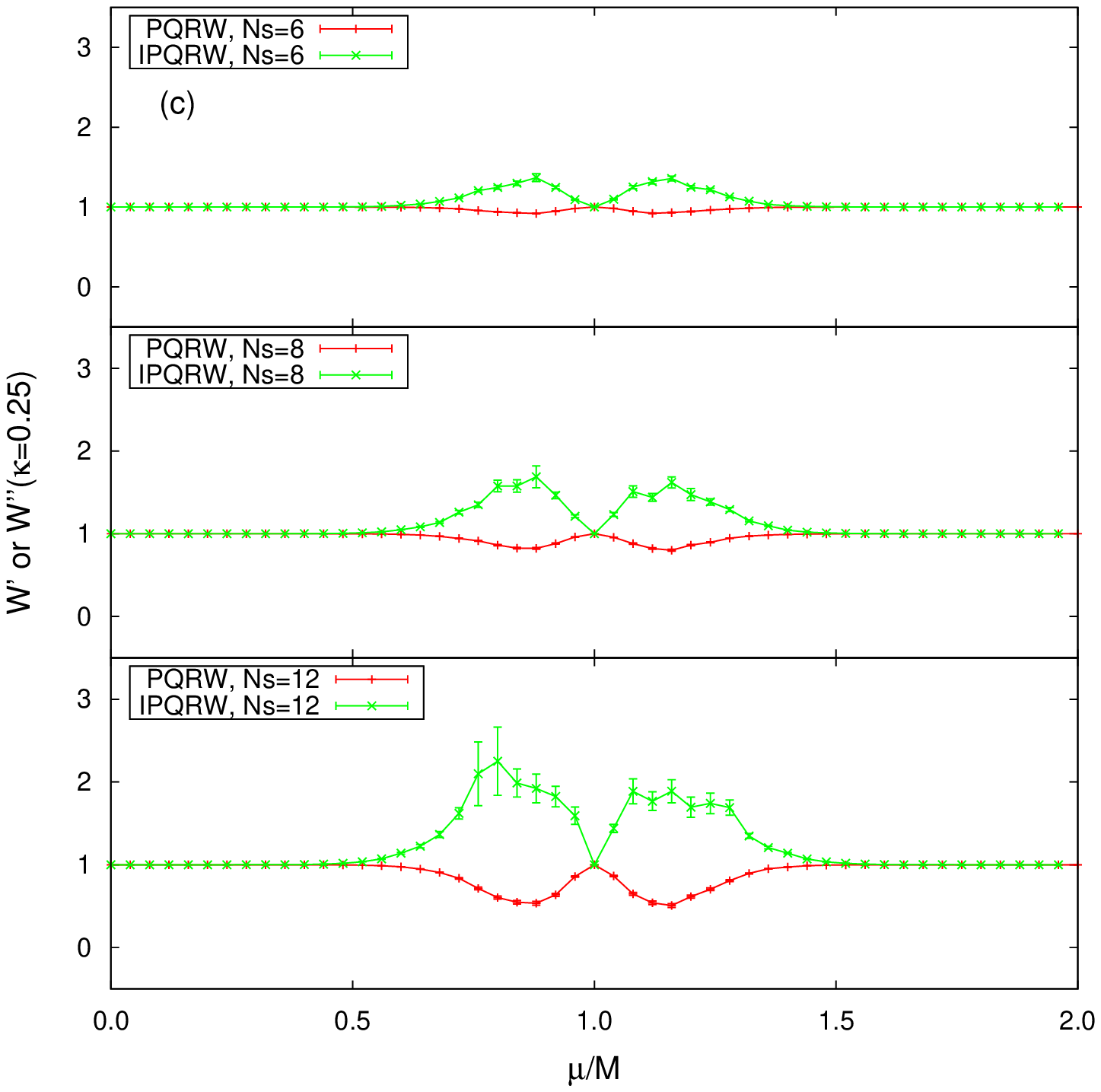}
\end{center}
\caption{The factors $W^\prime$ and $W^{\prime\prime}$ in EPLMWO at $M/T =10$, where PQRW and IPQRW are used. 
We set $\alpha =3.5$ in IPQRW. 
Note that $W^{\prime\prime}$ is not normalized in this figure. 
(a) $\kappa =0$, (b) $\kappa =0.09$, (c) $\kappa =0.25$. 
}
\label{pf_EPLM_k0_I}
\end{figure}
%%%%%%%%%%%%

%%%%%%%%%%%%%%%%%%%%%%
\begin{figure}[htbp]
\begin{center}
\includegraphics[width=0.4\textwidth]{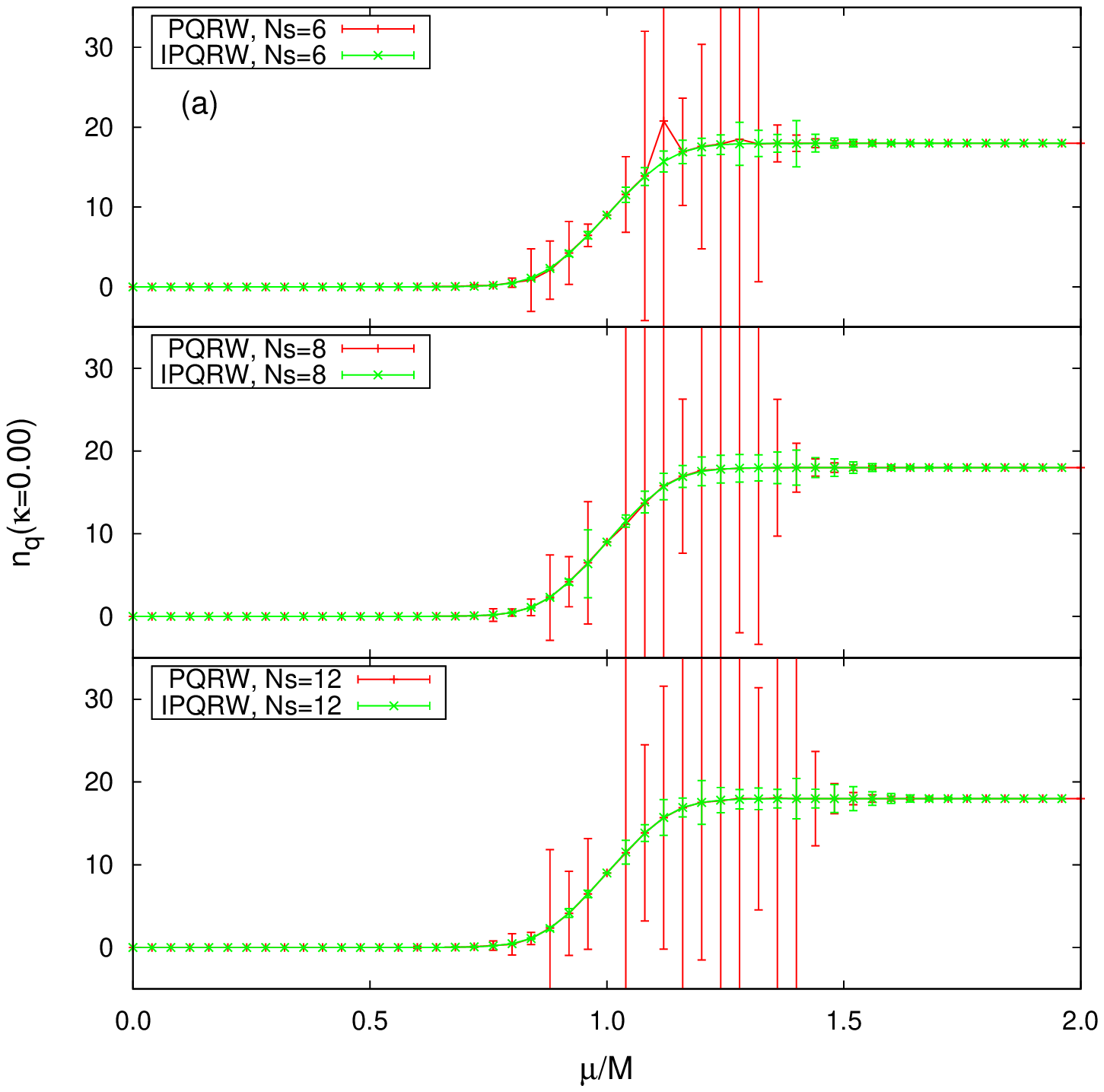}
\includegraphics[width=0.4\textwidth]{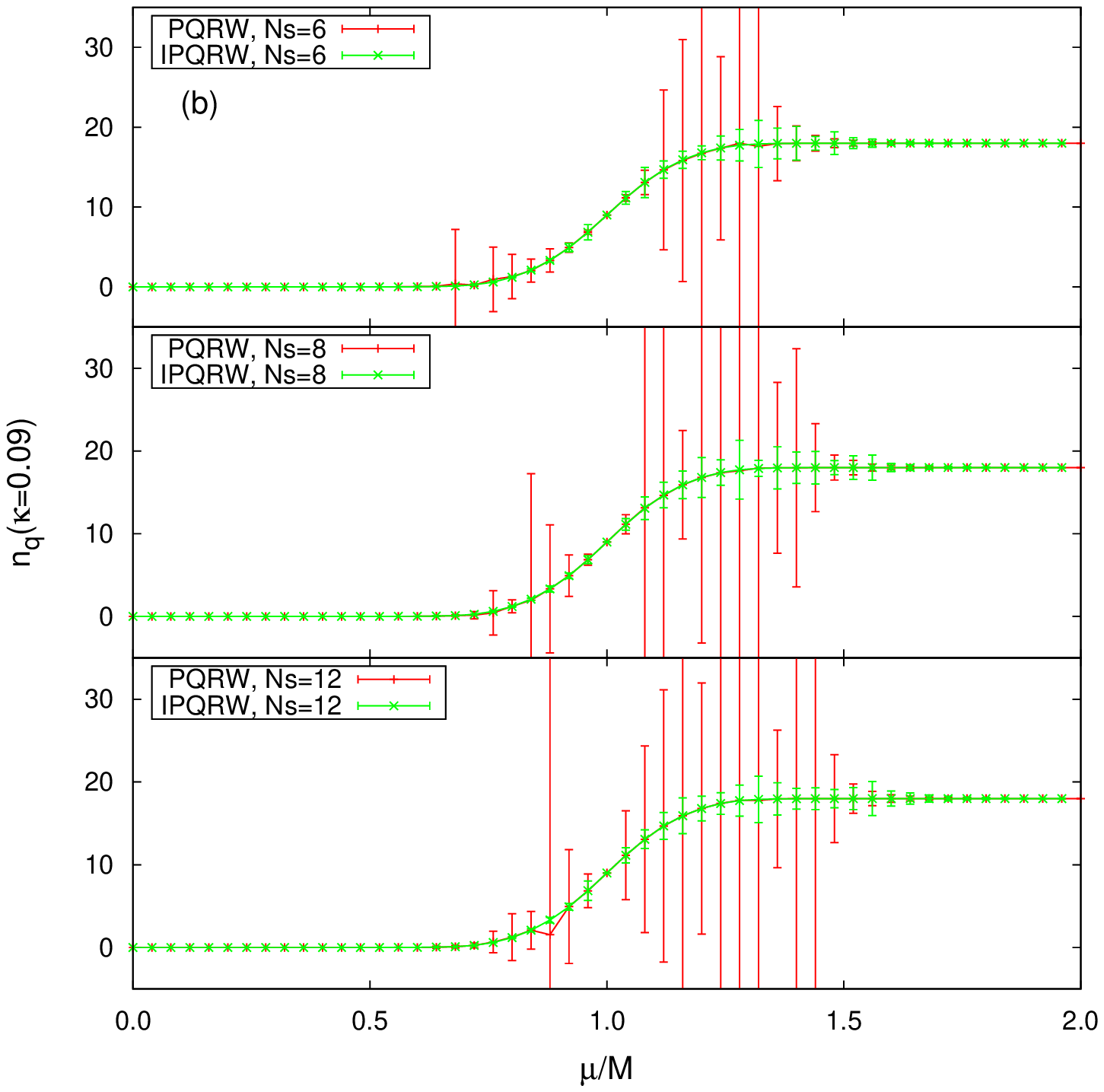}
\includegraphics[width=0.4\textwidth]{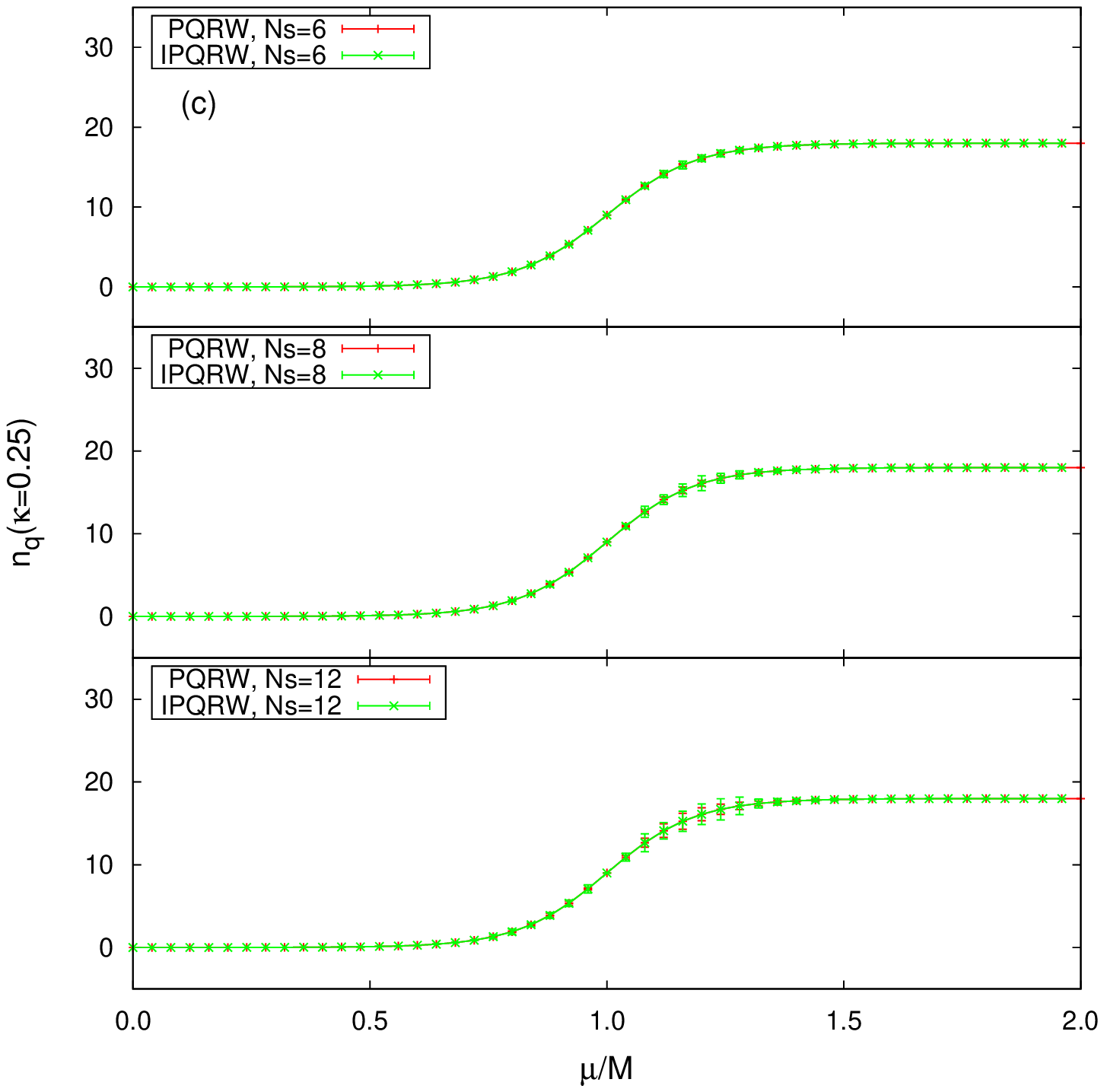}
\end{center}
\caption{The quark number density $n_q$ in EPLMWO at $M/T =10$, where PQRW and IPQRW are used. 
We set $\alpha =3.5$ in IPQRW. 
(a) $\kappa =0$, (b) $\kappa =0.09$, (c) $\kappa =0.25$.
}
\label{nq_EPLM_k0_I}
\end{figure}
%%%%%%%%%%%%

%%%%%%%%%%%%%%%%%%%%%%
\begin{figure}[htbp]
\begin{center}
\includegraphics[width=0.4\textwidth]{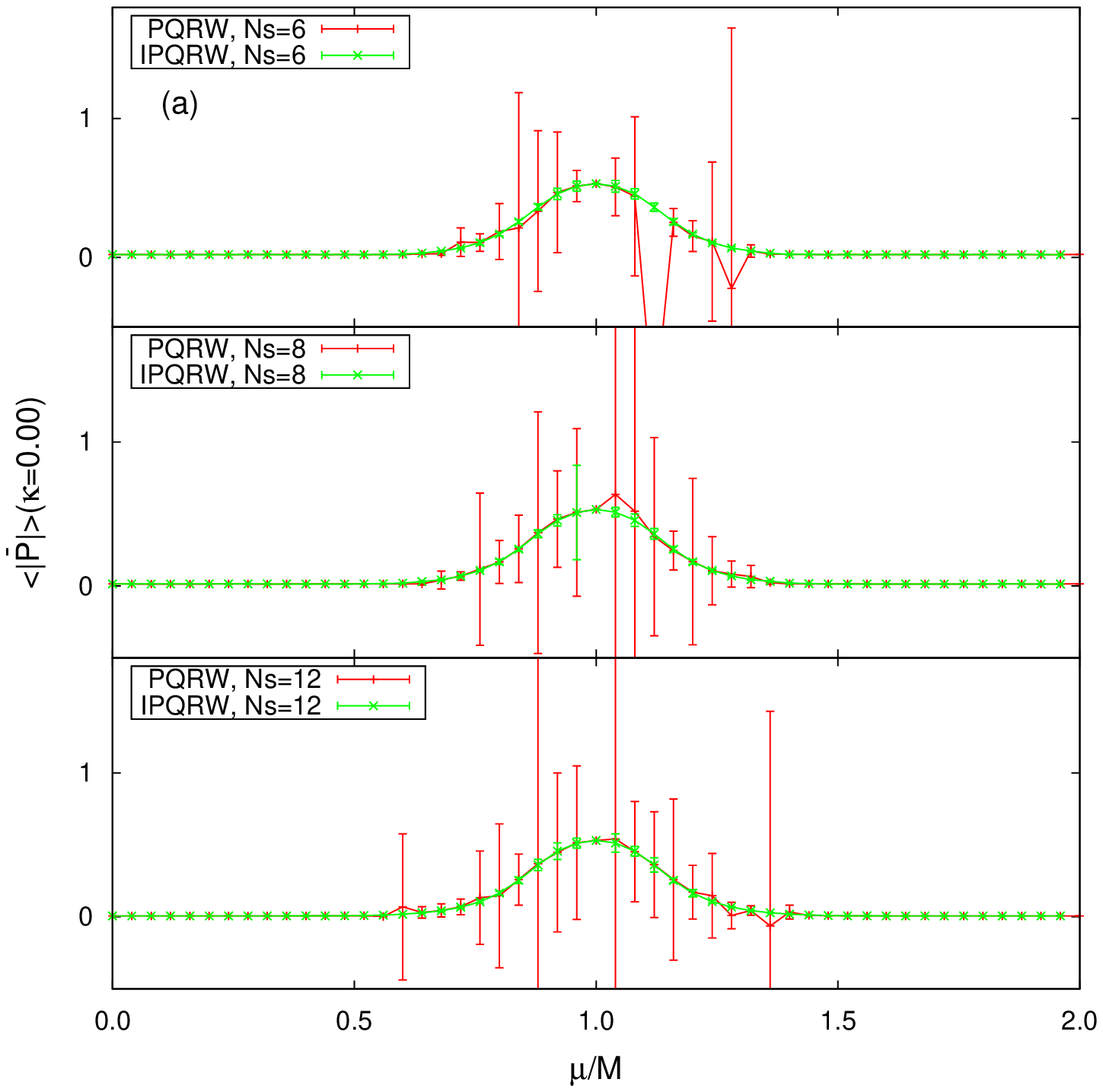}
\includegraphics[width=0.4\textwidth]{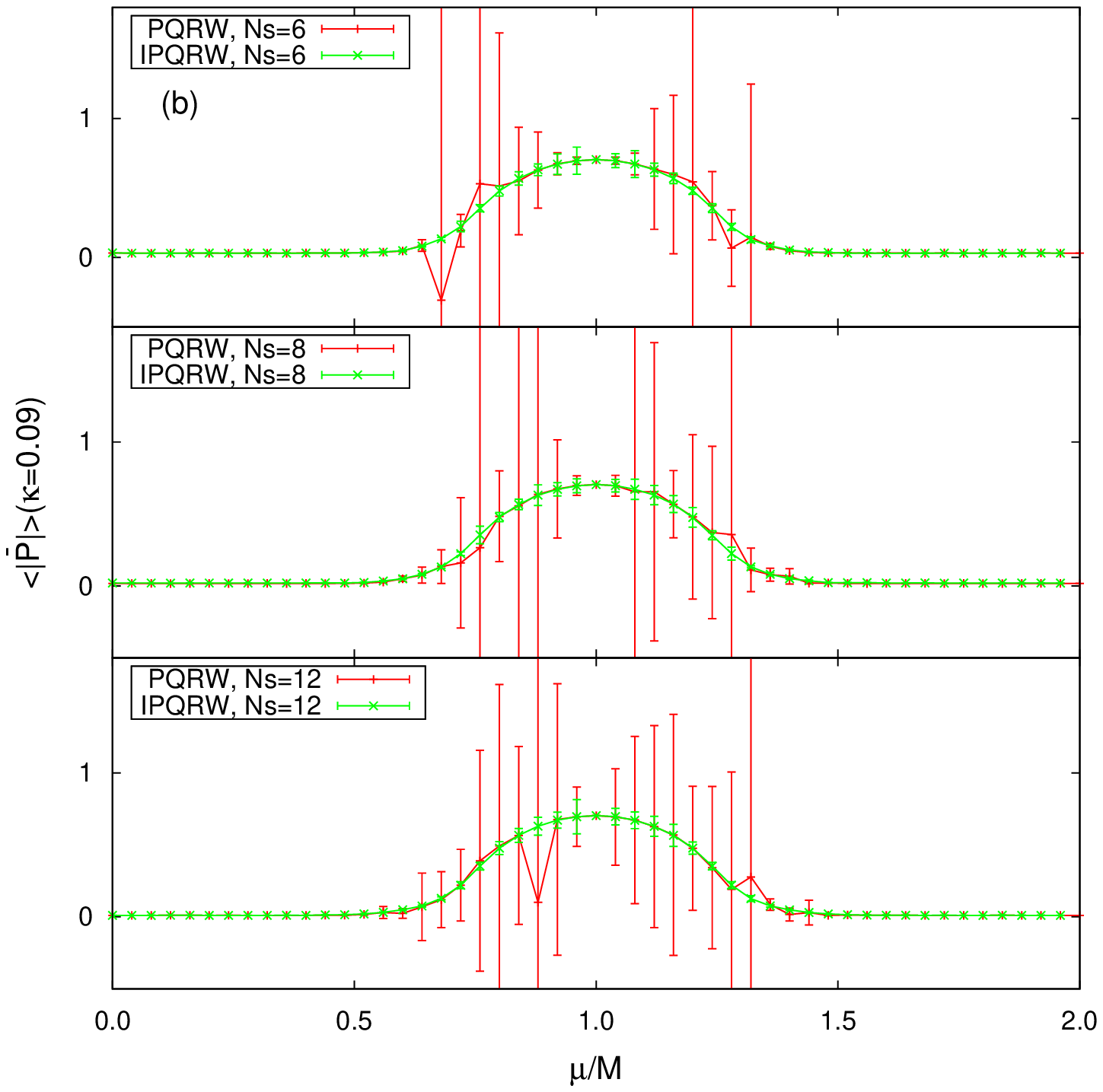}
\includegraphics[width=0.4\textwidth]{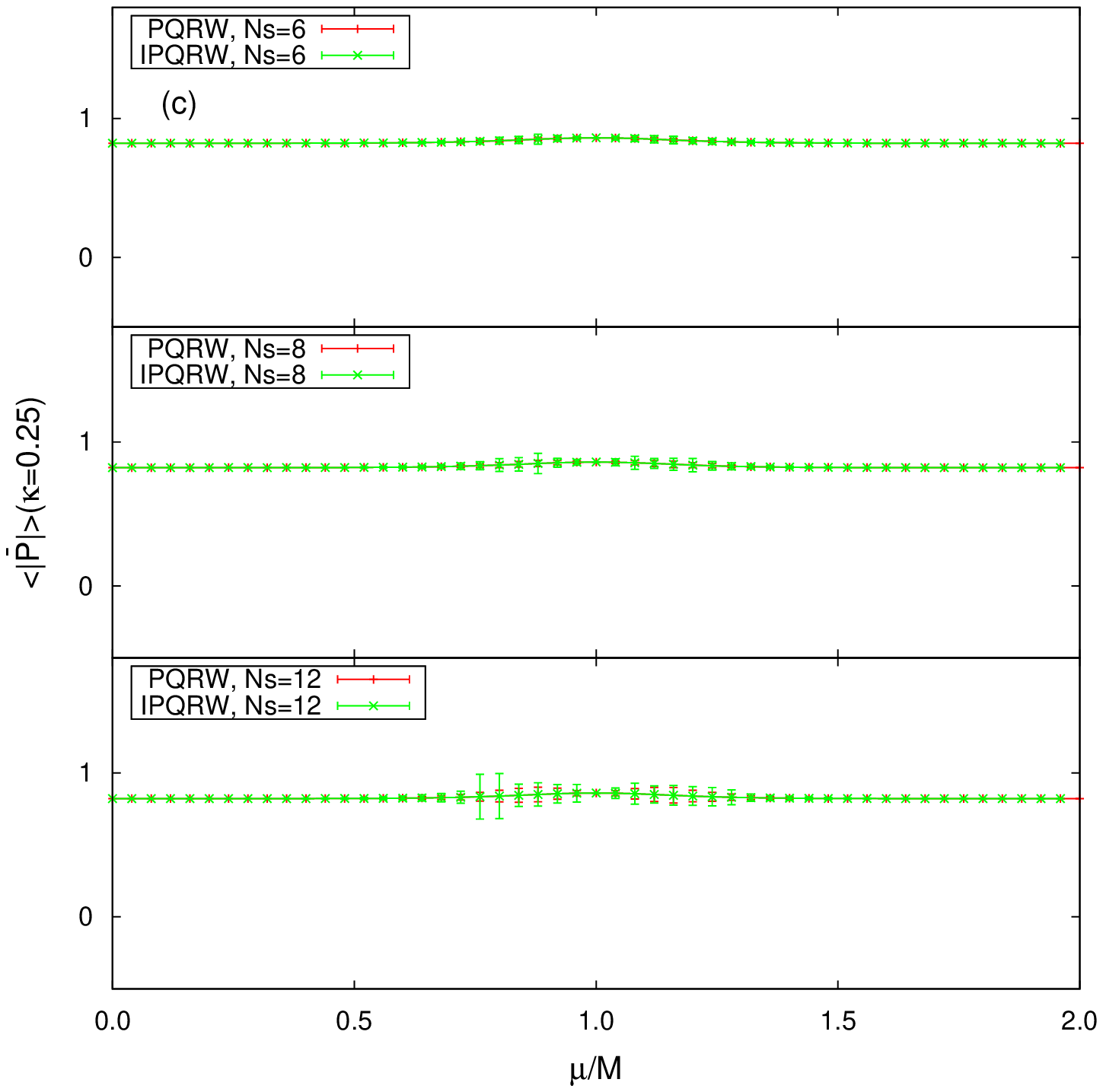}
\end{center}
\caption{The expectation value $\langle |\bar{P}|\rangle$ of the Polyakov line in EPLMWO at $\kappa =0$, $M/T =10$, where PQRW and IPQRW are used. 
We set $\alpha =3.5$ in IPQRW. 
(a) $\kappa =0$, (b) $\kappa =0.09$, (c) $\kappa =0.25$.}
\label{P_EPLM_k0_I}
\end{figure}
%%%%%%%%%%%%

Figure~\ref{pf_EPLM_I} shows $W^{\prime\prime}$ normalized by $\tilde{W}^{\prime\prime}$ for EPLMWO, where IPQRW is used. 
Note that the absolute value of the normalized $W^{\prime\prime}$ is not larger than the unnormalized one, since $\tilde{W}^{\prime\prime}\ge 1$ is ensured. 
Comparing it with the $W^\prime$ obtained by PQRW, we can find that 
the absolute value of the normalized $W^{\prime\prime}$ 
becomes somewhat large. 
However, it is still considerably small when $\mu /M=0.6\sim 0.14$ and $\kappa <0.12$ except for the line $\mu =M$.  

Figures~\ref{pf_EPLM_k0_I} shows 
the $\mu$ dependence 
of $W^\prime$ and $W^{\prime\prime}$ for EPLMWO. 
In the region $\mu /M=0.6\sim 1.4$, the ratio $W^\prime$ in PQRW is close to zero except for the case of $\mu =M$, when $\kappa =0$ and 0.09. 
In the same region, the ratio $W^{\prime\prime}$ in IPQRW lies between 1 and 2 but it fluctuates to some extent. 
 
Figures~\ref{nq_EPLM_k0_I} shows the $\mu$ dependence of $n_q$ for EPLMWO. 
In the region $\mu /M=0.6\sim 1.4$ with $\kappa =0$ and 0.09, due to the smallness of $W^\prime$, the density $n_q$ has a large error when PQRW is used, except for the case of $\mu =M$. 
When IPQRW is used, 
the error of $n_q$ are small in the same region. 
In the figure, it is also seen that the $N_s$ dependence is small when IPQRW is used. 

Figures~\ref{P_EPLM_k0_I} shows the $\mu$ dependence of 
$\langle |\bar{P}|\rangle$ for EPLMWO. 
As in the case of $n_q$, in the region $\mu /M=0.6\sim 1.4$ with $\kappa =0$ and 0.09, due to the smallness of $W^\prime$, 
the expectation value $\langle |\bar{P}|\rangle$ has a large error when PQRW is used, except for the case of $\mu =M$. 
When IPQRW is used, the error of $\langle |\bar{P}|\rangle$ are small. 
Again, the $N_s$ dependence is small, when IPQRW is used.

%%%%%%%%%%%%%%%%%%%%%%
\begin{figure}[htbp]
\begin{center}
\includegraphics[width=0.4\textwidth]{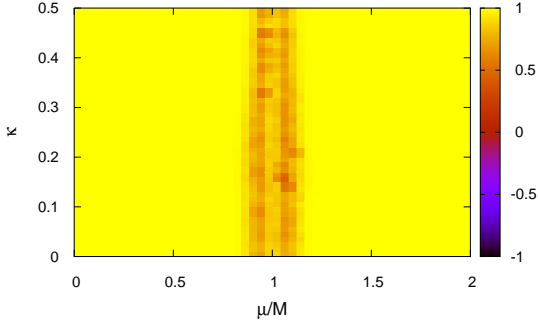}
\end{center}
\caption{The ratio $W^{\prime\prime}/\tilde{W}^{\prime\prime}$ in $Z_3$-EPLM, where IPQRW is used. 
We set $\alpha =3.5$ and $N_s=6$.  
}
\label{Fig_Phase_factor_Z3_I}
\end{figure}
%%%%%%%%%%%%

%%%%%%%%%%%%%%%%%%%%%%
\begin{figure}[htbp]
\begin{center}
\includegraphics[width=0.4\textwidth]{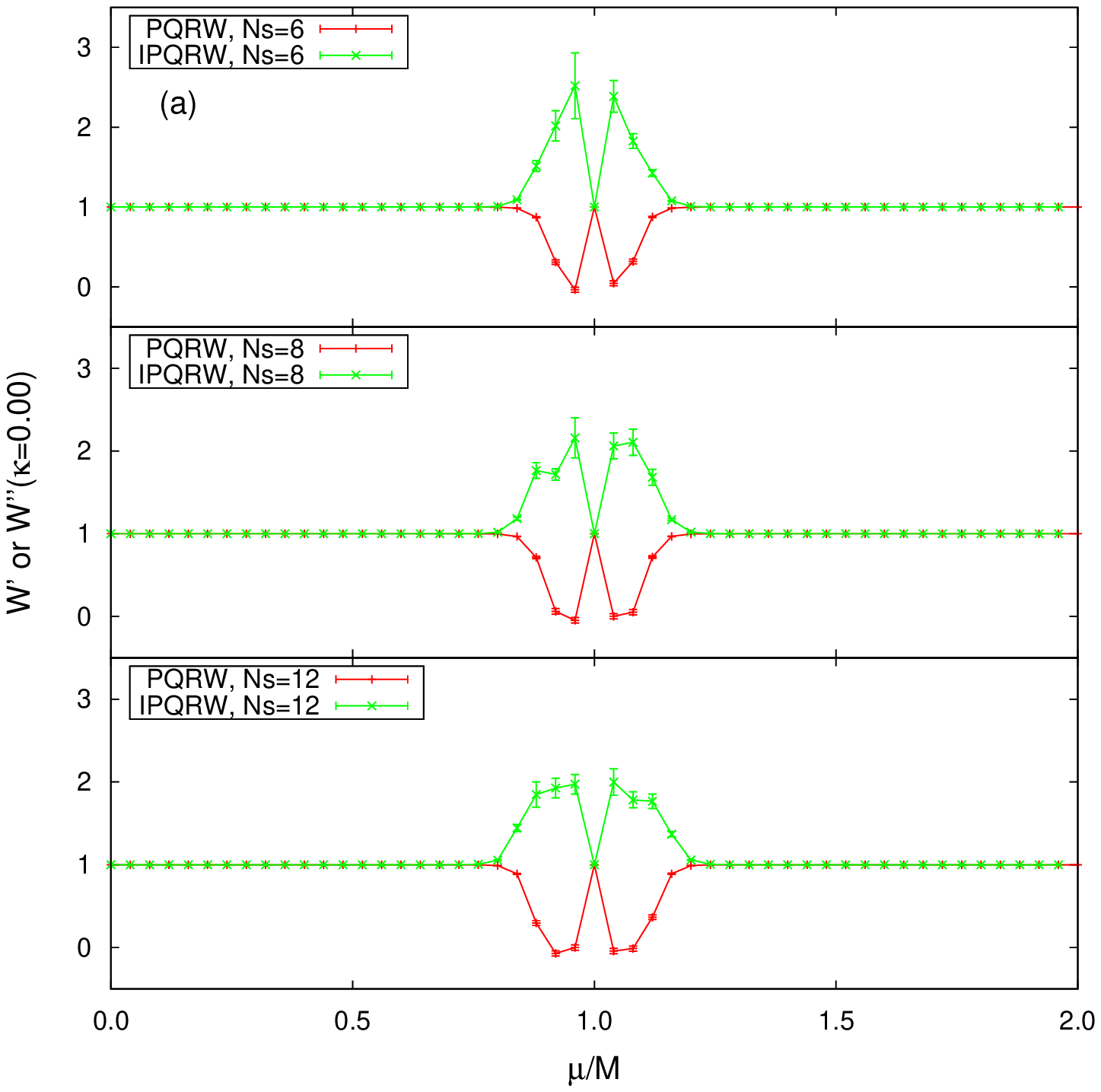}
\includegraphics[width=0.4\textwidth]{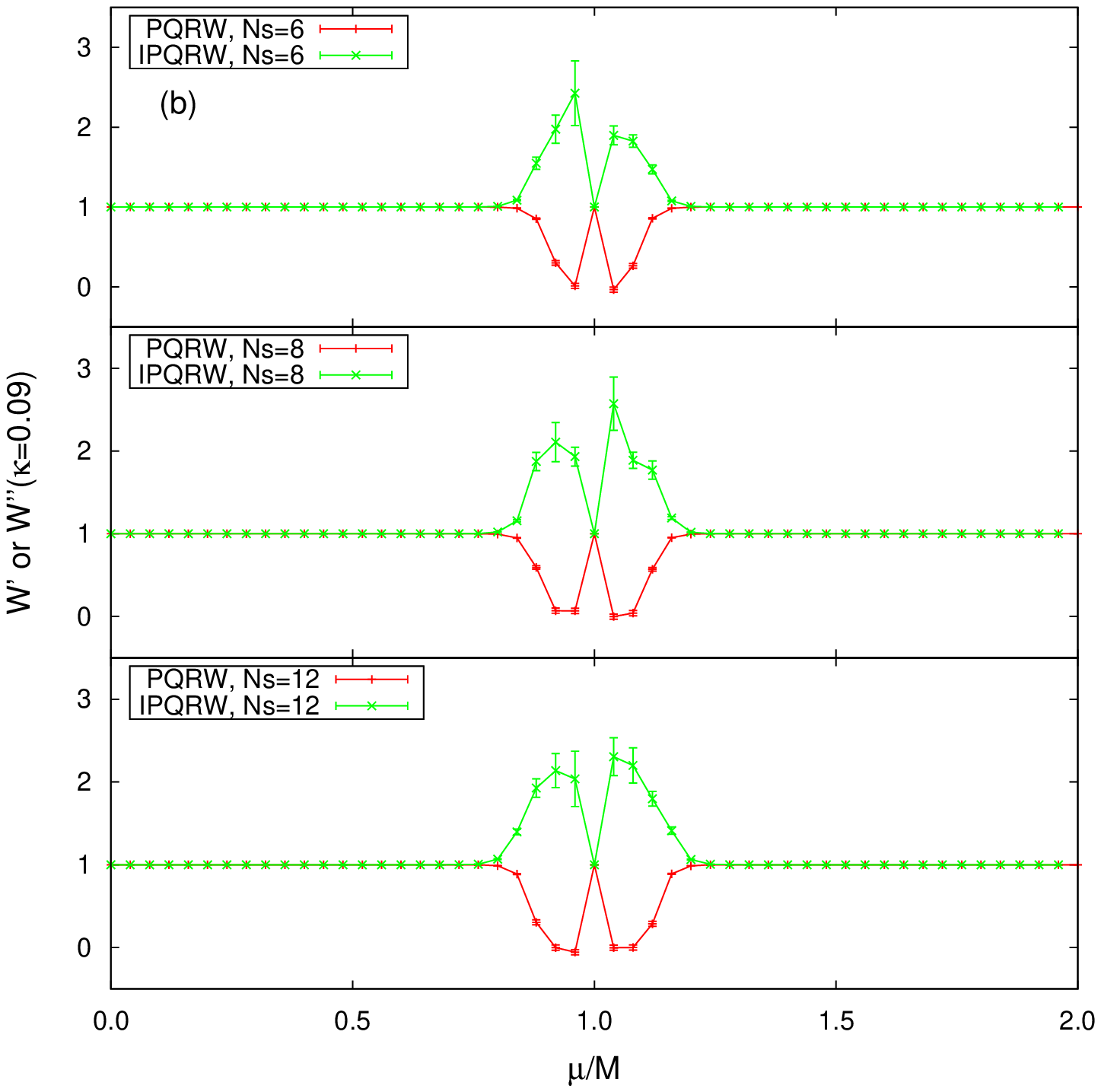}
\includegraphics[width=0.4\textwidth]{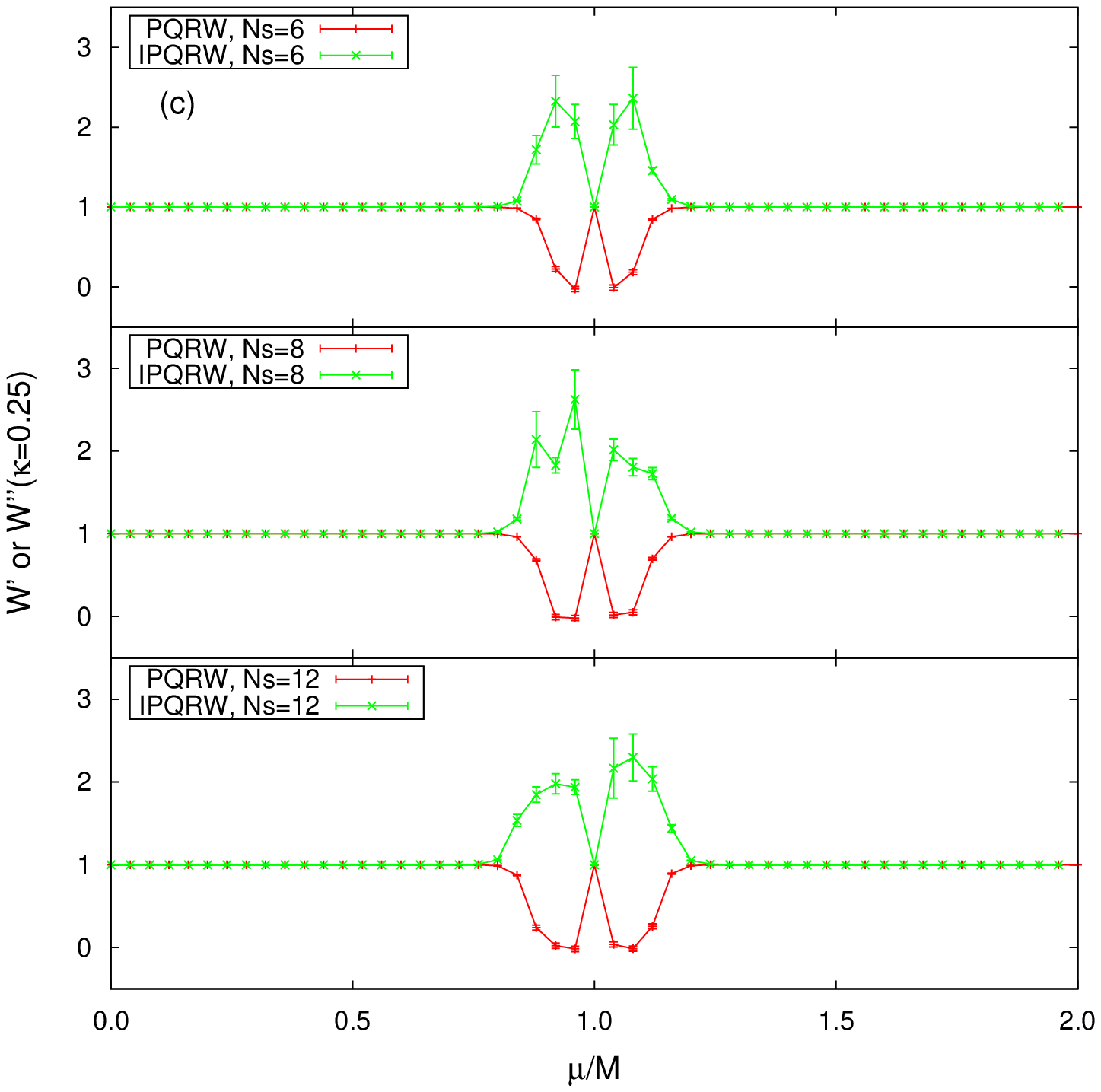}
\end{center}
\caption{The factors $W^{\prime}$ and $W^{\prime\prime}$ in $Z_3$-EPLM at $\kappa =0$, $M/T =10$, where PQRW and IPQRW are used. 
We set $\alpha =3.5$ in IPQRW. 
Note that $W^{\prime\prime}$ is not normalized in this figure. 
(a) $\kappa =0$, (b) $\kappa =0.09$, (c) $\kappa =0.25$.}
\label{Fig_rf_Z3_M10_IPQRW}
\end{figure}
%%%%%%%%%%%%

%%%%%%%%%%%%%%%%%%%%%%
\begin{figure}[htbp]
\begin{center}
\includegraphics[width=0.4\textwidth]{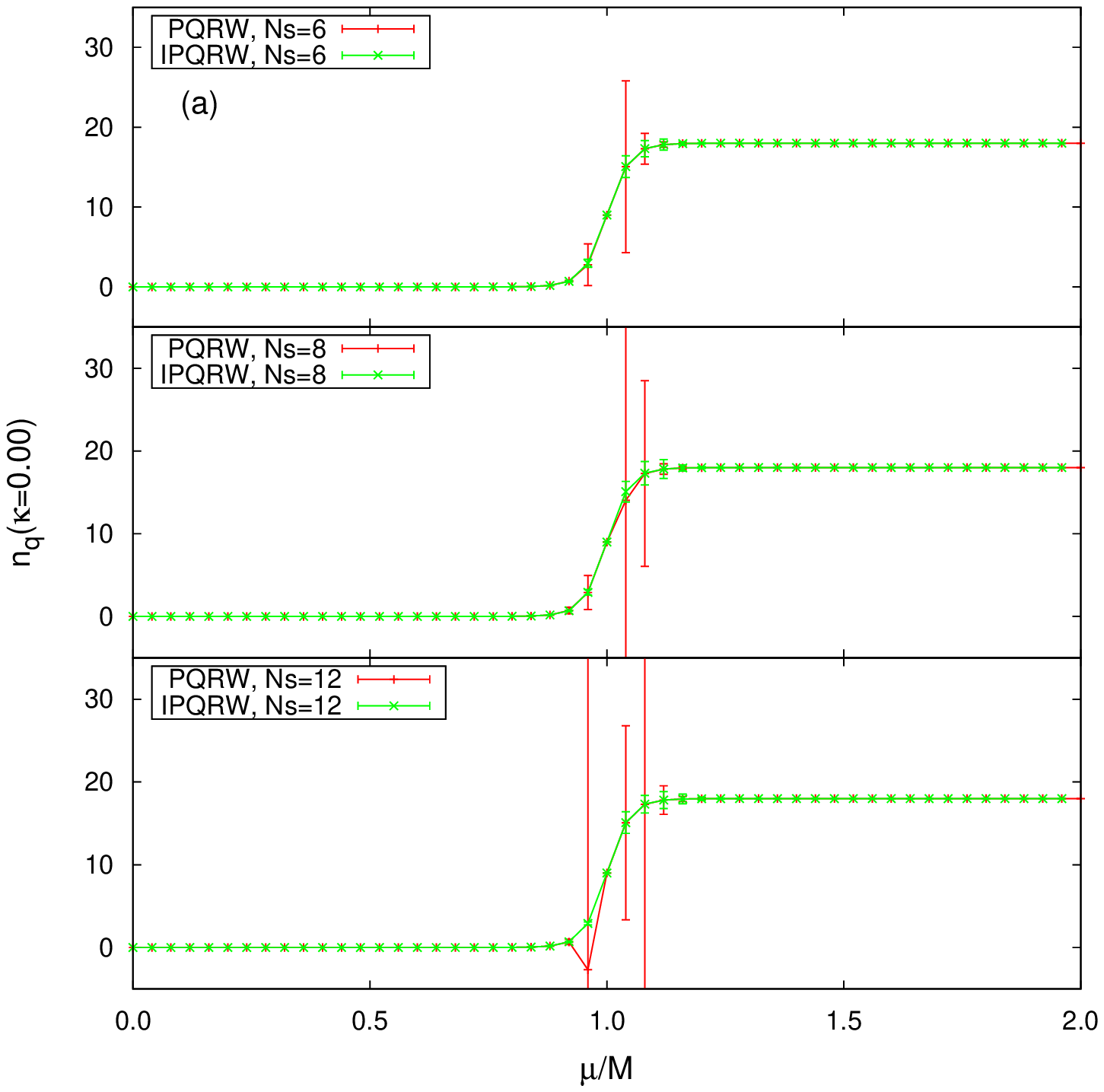}
\includegraphics[width=0.4\textwidth]{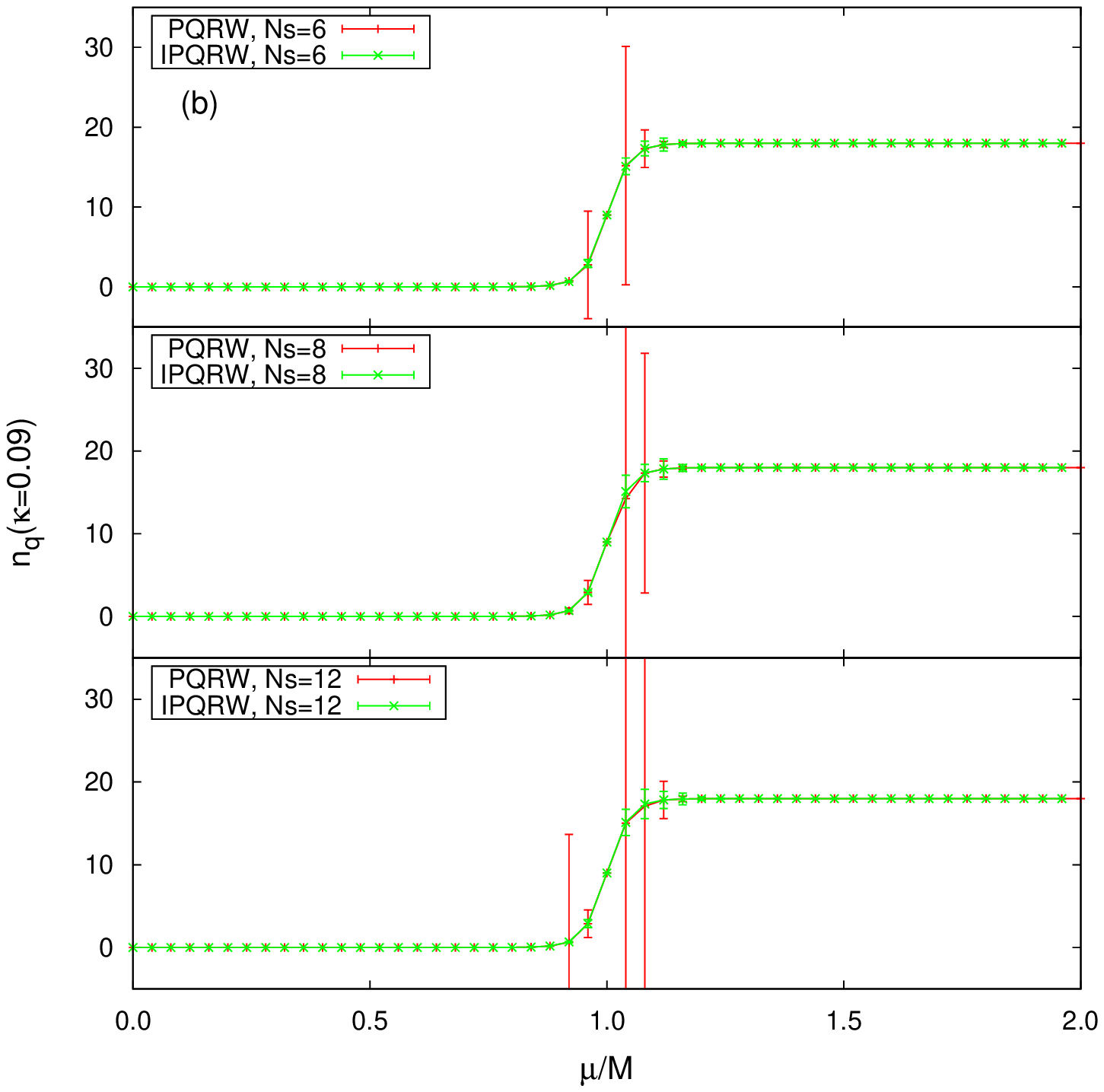}
\includegraphics[width=0.4\textwidth]{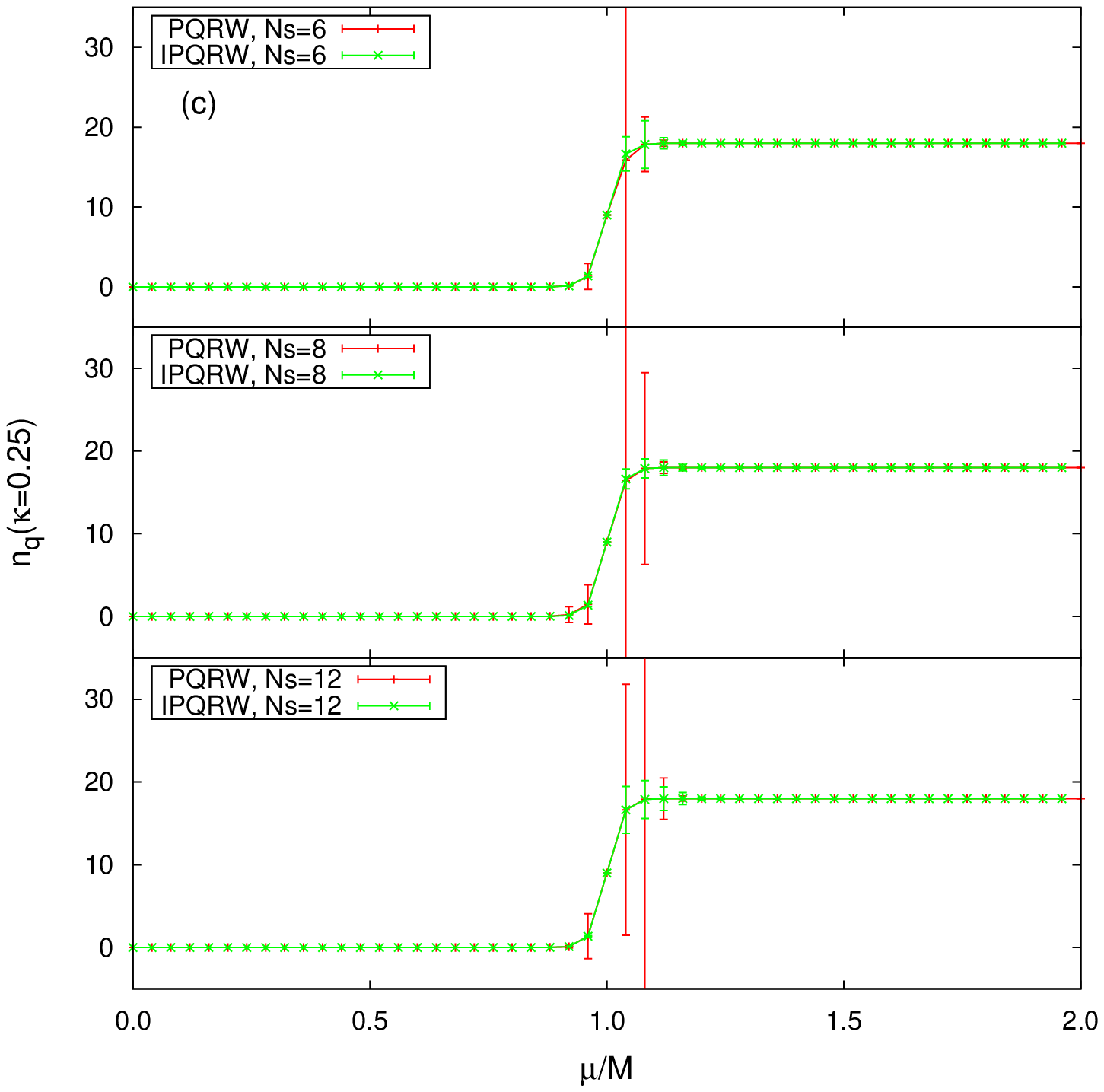}
\end{center}
\caption{The quark number density $n_q$ in $Z_3$-EPLM at $\kappa =0$, $M/T =10$, where PQRW and IPQRW are used. 
We set $\alpha =3.5$ in IPQRW. 
(a) $\kappa =0$, (b) $\kappa =0.09$, (c) $\kappa =0.25$.
}
\label{Fig_nq_Z3_M10_IPQRW}
\end{figure}
%%%%%%%%%%%%

%%%%%%%%%%%%%%%%%%%%%%
\begin{figure}[htbp]
\begin{center}
\includegraphics[width=0.4\textwidth]{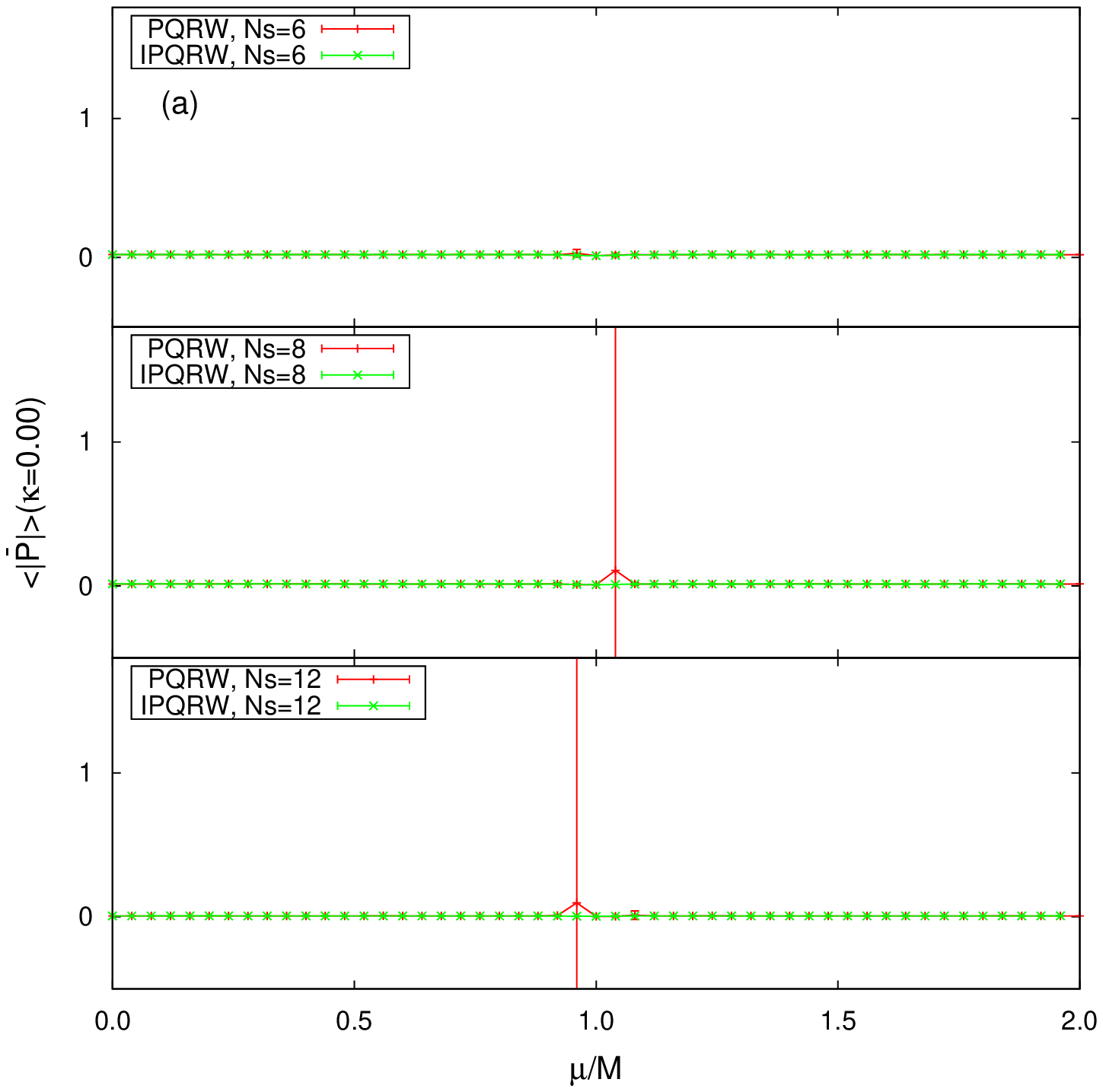}
\includegraphics[width=0.4\textwidth]{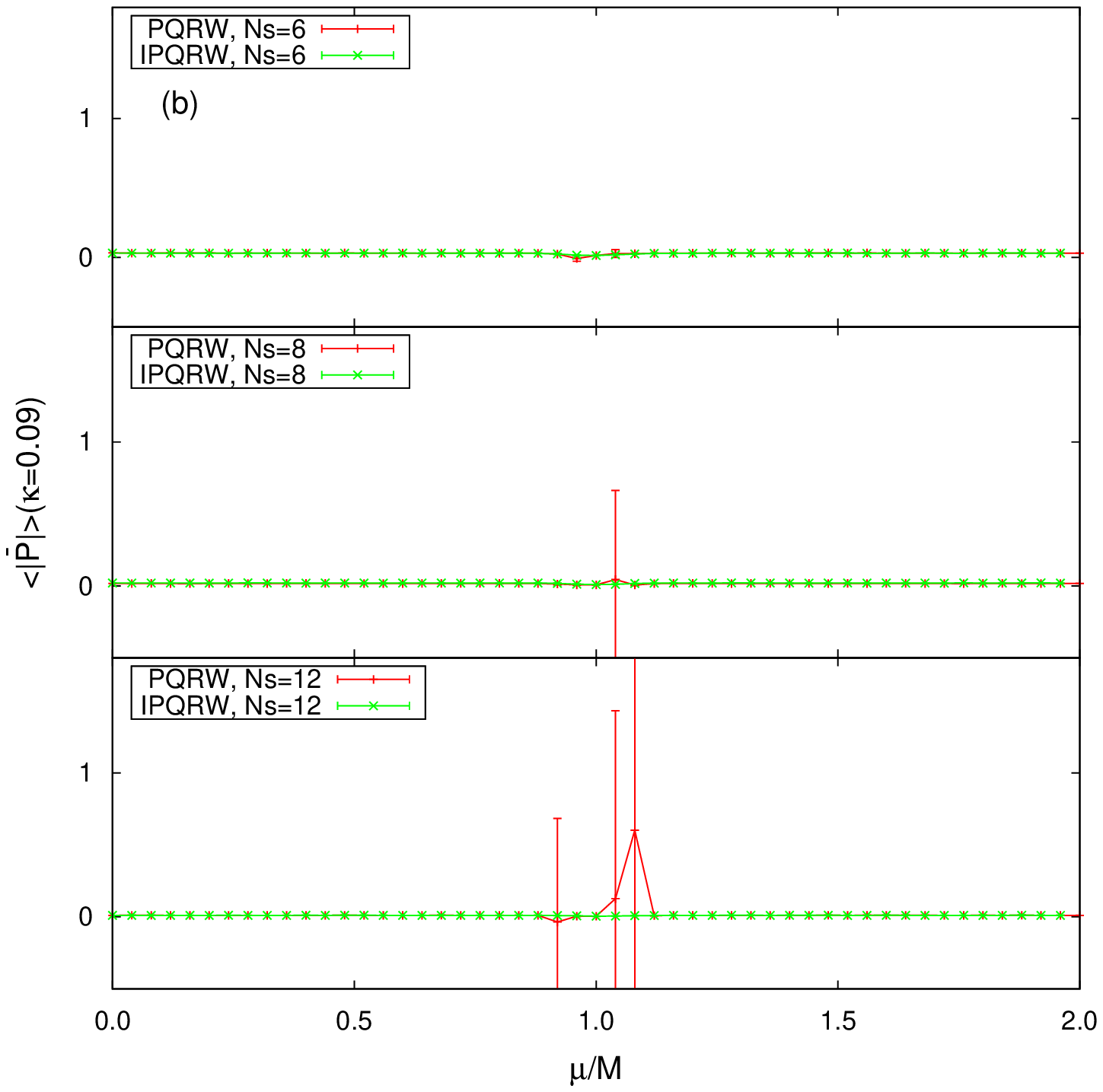}
\includegraphics[width=0.4\textwidth]{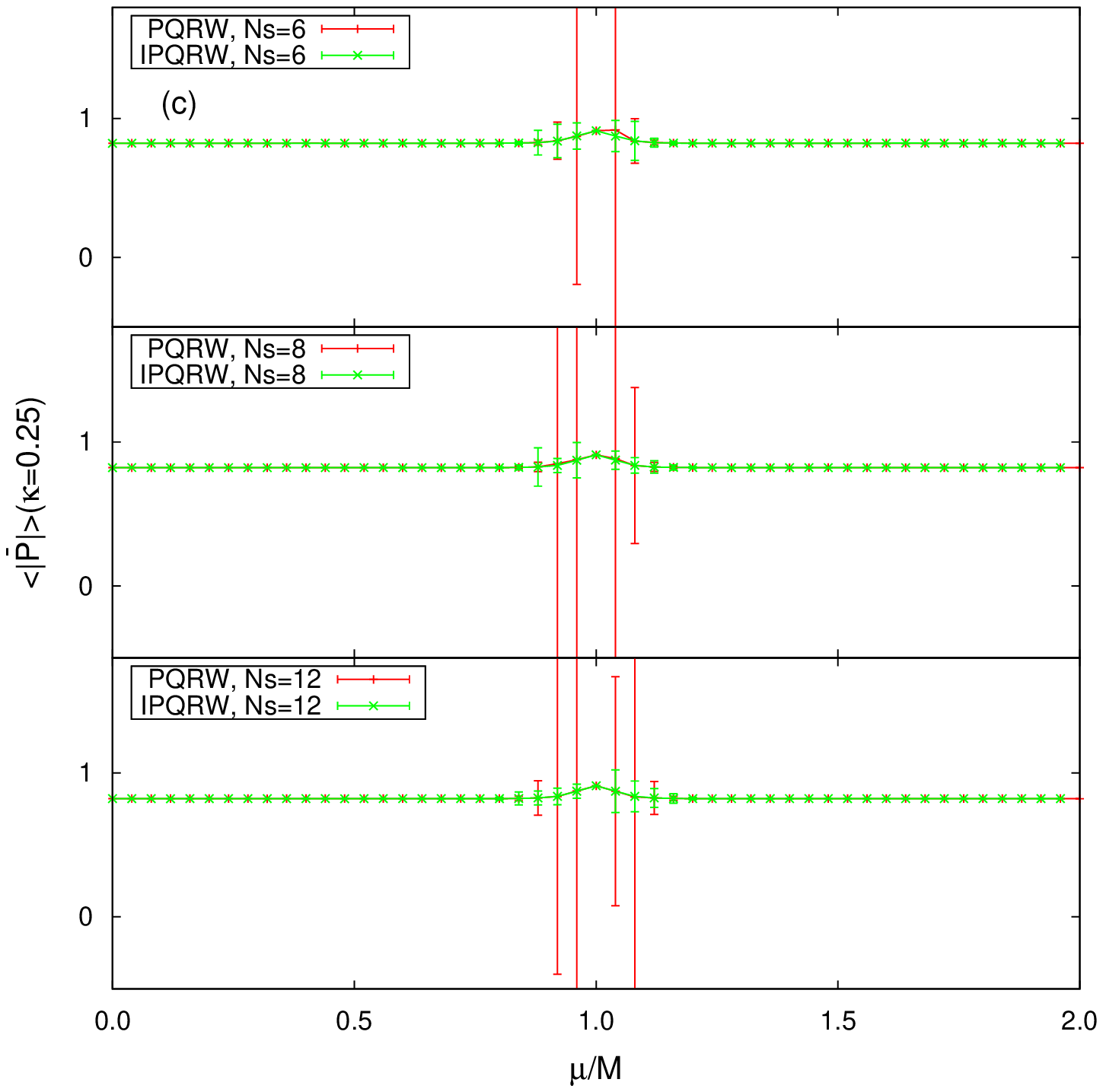}
\end{center}
\caption{The expectation value $\langle \bar{P}\rangle$ of the Polyakov-line in $Z_3$-EPLM at $\kappa =0$, $M/T =10$, where PQRW and IPQRW are used. 
We set $\alpha =3.5$ in IPQRW. 
(a) $\kappa =0$, (b) $\kappa =0.09$, (c) $\kappa =0.25$.}
\label{Fig_P_Z3_M10_IPQRW}
\end{figure}
%%%%%%%%%%%%

Figure~\ref{Fig_Phase_factor_Z3_I} shows  the $W^{\prime\prime}$ normalized 
by $\tilde{W}^{\prime\prime}$ for $Z_3$-EPLM, where IPQRW is used.   
The normalized $W^{\prime\prime}$ is close to 1 even in the vicinity of $\mu=M$.   
Even for small $\kappa$, the sign problem almost vanishes.  

Figure~\ref{Fig_rf_Z3_M10_IPQRW} shows the $\mu$ dependence of 
$W^\prime$ and $W^{\prime\prime}$ for  $Z_3$-EPLM. 
For all the cases of $\kappa$, in the vicinities of $\mu /M=0.95$ or $1.05$, the ratio $W^\prime$ in PQRW is close to zero.  
In the same region, the ratio factor $W^{\prime\prime}$ in IPQRW is close to 2. 

Figure~\ref{Fig_nq_Z3_M10_IPQRW} shows the $\mu$ dependence of $n_q$ 
for $Z_3$-EPLM. 
In the vicinity of $\mu /M=0.95$ and also of $\mu /M=1.05$, 
due to the smallness of $W^\prime$, the density $n_q$ has a large error except for the case of $\mu =M$ when PQRW is used. 
When IPQRW is used, the error of $n_q$ is small.  
It is also seen that the $N_s$ dependence is small, when IPQRW is used. 

Figure~\ref{Fig_P_Z3_M10_IPQRW} shows the $\mu$ dependence of $\langle |\bar{P}|\rangle$ for $Z_3$-EPLM. 
For $\kappa =0$ and 0.09, the mean value of $\langle |\bar{P}|\rangle$ almost vanishes for any $\mu$. 
In the vicinity of $\mu /M=0.95$ and also of $\mu /M=1.05$, due to the smallness of $W^\prime$, the expectation value 
$\langle |\bar{P}|\rangle$ has a large error when PQRW is used, except for the case with $\kappa =0 $ or 0.09 and $N_s=6$, where the numerator of the last line in Eq. (\ref{rw_phase_PQA})  vanishes almost completely.  
When IPQRW is used, the error of $\langle |\bar{P}|\rangle$ are small. 
The $N_s$ dependence is small in the figure when IPQRW is used. 
 
%%%%%%%%%%%%%%%%%%%%%%
\begin{figure}[htbp]
\begin{center}
\includegraphics[width=0.4\textwidth]{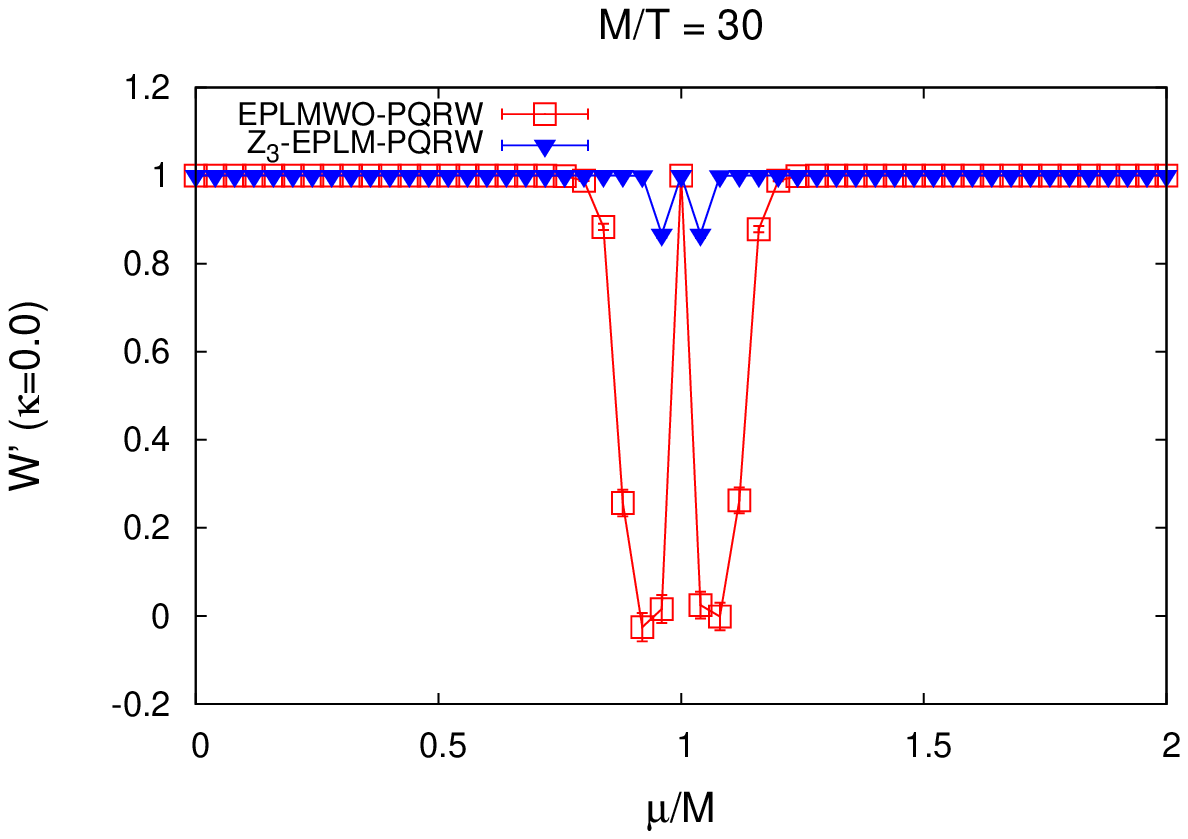}
\end{center}
\caption{The phase factors in EPLMWO and $Z_3$-EPLM at $\kappa =0$, $M/T =30$, where PQRW is used. }
\label{Fig_rf_M30}
\end{figure}
%%%%%%%%%%%%

%%%%%%%%%%%%%%%%%%%%%%
\begin{figure}[htbp]
\begin{center}
\includegraphics[width=0.4\textwidth]{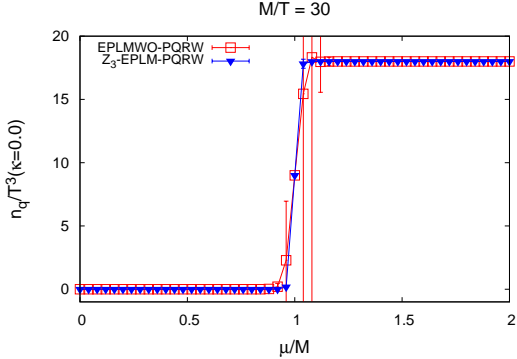}
\end{center}
\caption{The quark number density $n_q$ in EPLMWO and $Z_3$-EPLM at $\kappa =0$, $M/T =30$, where PQRW is used. }
\label{Fig_nq_M30}
\end{figure}
%%%%%%%%%%%%

In the calculations mentioned above, the $\mu$ dependence of $n_q$ at $\kappa =0$ in $Z_3$-EPLM is somewhat different from in EPLMWO. 
This result seems to be inconsistent with the expectation that $Z_3$-EPLM tends to EPLMWO when $T\to 0$. 
However, in the calculations shown above, we have taken the case of $M/T=10$. 
In the limit $T\to 0$, we should consider the limit $M/T\to \infty$ at the same time to put $\kappa =0$. 
In Figs. \ref{Fig_rf_M30} and \ref{Fig_nq_M30}, the $\mu$ dependences of the phase factor and $n_q$ at $M/T=30$ and $\kappa =0$ are shown 
in both EPLMWO and $Z_3$-EPLM, where PQRW is used. 
We see that the sign problem is very weak everywhere in $Z_3$-EPLM while it is 
still strong at $\mu /M\sim 0.95$ in EPLMWO. 
The quark number density $n_q$ in EPLMWO almost coincides with that in $Z_3$-EPLM when the sign problem is weak in EPLMWO. 
This result indicates that $Z_3$-EPLM tends to EPLMWO in the limit $T\to 0$.  
We also see that $n_q(\mu )$ is close to $18\Theta(\mu-M)$ in $Z_3$-EPLM, where $\Theta(x)$ is a step function and $18$ is the degree of freedom of quark.
This suggests that the early onset of $n_q$ does not appear at low $T$ for 
the case of $Z_3$-EPLM.  

%%%%%%%%%%%%%%%%%%%%%%%%%%%%%%%%%%%%%%%%%%%%%%%%%%%%%%%%
%%%% summary
%%%%%%%%%%%%%%%%%%%%%%%%%%%%%%%%%%%%%%%%%%%%%%%%%%%%%%%%
\section{Summary}
\label{summary}
%%%%%%%%%%%%%%%%%%%%%%%%%%%%%%%%%%%%%%%%%%%%%%%%%%%%%%%%

In summary, we have studied the sign problem in the $Z_3$-symmetric 
effective Polyakov-line model ($Z_3$-EPLM). 
In $Z_3$-EPLM, the confinement state of $P_\vix=0$ 
and the three deconfinement states based on $Z_3$ symmetry degenerate, 
when the pure gauge contribution is neglected. 
In other words, the pure gauge contribution is important to determine which state is more favorable. 
The Haar measure term favors the confinement state, while the gluonic kinetic 
term does the  ordered deconfinement configurations. 

In the confinement phase where the effect of gluonic kinetic term is small, the confinement state of $P_\vix=0$ is favored but finite $P_\vix$ is also realized to some extent as the fluctuations. 
However, the realized probability $P_\vix$ is almost $Z_3$-symmetric even in one configuration, and consequently, the spatial average $\bar{P}=\sum_\vix P_\vix$ almost vanishes. 
Meanwhile, in the deconfinement phase, the ordered configuration is favored and the spatial average $\bar{P}$ is finite in one configuration. 
Due to $Z_3$ symmetry, the configuration average $\langle \bar{P}\rangle$ vanishes but $\langle |\bar{P}| \rangle$ does not. 

$Z_3$-EPLM has no sign problem, when either 
the confinement state or the three deconfinement states are realized. 
This is  because $P_\vix$ is real in 
both the confinement and deconfinement states. 
In fact, this happens in the case of $Z_3$-symmetric 3 or 4 states Potts model~\cite{Hirakida}. 

In both the random confinement phase and the intermediate phase, however, $P_\vix$ fluctuates considerably and has finite imaginary part. 
Hence, ${\rm Im}[S_{\rm F}]$ is finite for finite $\mu$. 
This causes a sign problem. 
Nevertheless, due to the reality of the confinement and the deconfinement states mentioned above and the smallness of ${\rm Im}[S_{\rm F}]$, the sign problem is expected to be considerably milder in $Z_3$-EPLM than in the ordinary EPLM with no $Z_3$-symmetry.  
In fact, the results obtained by the phase quenched reweighting method show that the sign problem is considerably milder in $Z_3$-EPLM than in the ordinary EPLM with no $Z_3$-symmetry when $\kappa$ is small.    

We have also proposed the new reweighting method to include the contribution of the imaginary part of the fermion effective action into the approximate distribution function. 
The new method makes the sign problem somewhat milder.  
Particularly for in $Z_3$-EPLM, 
the sign problem is very weak at small $\kappa$. 
It is also found that the results depend on $N_s$ only weakly 
when the improved reweighting method is used.  

Our results also indicate that the early onset of quark number density $n_q$ does not appear in $Z_3$-EPLM, when $T$ is small. 
However, it may not be the case of $Z_3$-QCD.  
The effective Polyakov-line model has the dynamics related to the Polyakov-line but is not expected to include the chiral dynamics in the calculation, since 
the quark mass is very large. 
In LQCD calculations, the pion-condensation-like phenomena appear as an artifact for finite $\mu$, when the phase quenched reweighting is used. 
This artifact induces the problem on the early onset of quark number density~\cite{Barbour} or the baryon-number Silver Blaze problem~\cite{Cohen}.  
In the improved reweighting method proposed in this paper, contributions of the imaginary part of the fermion action are included into the approximated 
probability function to some extent. 
This improvement may avoid the artifact mentioned above. 
It is very interesting to check whether this new method works well or not 
in LQCD simulations, particularly in lattice $Z_3$-QCD.

\noindent
\begin{acknowledgments}
The authors are thankful to Atsushi Nakamura, Hiroshi Yoneyama, Hiroshi Suzuki, Tatsuhiro Misumi, Etsuko Itou, Masahiro Ishii, Akihisa Miyahara, Shuichi Togawa, Yuhei Torigoe and K. Kashiwa for fruitful discussions. 
H. K. also thanks Masahiro Imachi, Hajime Aoki, Motoi Tachibana and Takahiro Doi for useful discussions. 
J. S., H. K. and M. Y. are/were supported by Grant-in-Aid for Scientific Research (No.27-7804, No.17K05446 and No. 26400279, and No.26400278) from Japan Society for the Promotion of Science (JSPS). 
 The numerical calculations were partially performed by using SX-ACE at Cybermedia Center and at RCNP, Osaka University.
 \\~
 \\~
 \\~
\end{acknowledgments}

\newpage

%%%%%%%%%%%%%%%%%%%%%%%%%%%%%%%%%%%%%%%%%%%%%%%%%%%%%%%%%%%%%%%%%%%%%%%%%%%%%%%%%%%%% References
%%%%%%%%%%%%%%%%%%%%%%%%%%%%%%%%%%%%%%%%%%%%%%%%%%%%%%%%%%%%%%%%%%%%%%%%%%%%%%%%

\end{document}